\documentclass[preprintnumbers,nofootinbib,amsmath,amssymb]{revtex4}

\def\ga{\mathrel{\raise.3ex\hbox{$>$\kern-.75em\lower1ex\hbox{$\sim$}}}}
\def\la{\mathrel{\raise.3ex\hbox{$<$\kern-.75em\lower1ex\hbox{$\sim$}}}}

\newcommand\beq{\begin{equation}}
\newcommand\eeq{\end{equation}}
\newcommand\beqar{\begin{eqnarray}}
\newcommand\eeqar{\end{eqnarray}}
\usepackage{graphicx}% Include figure files
\usepackage{epsfig}
\usepackage{epsf}
\usepackage{rotating}
\usepackage{verbatim}

 \usepackage{multirow}

\begin{document}

%\begin{titlepage}
%\pagestyle{empty}
%\baselineskip=21pt
%\vspace*{-0.6in}
\preprint{ArXiv:0812.1231} \preprint{UMN--TH--2728/08} \vskip 0.2in
%\begin{center}
\title{
Instability of the ACW model, and
problems with massive vectors during inflation
}
%\end{center}
%\begin{center}
%\vskip 0.2in
%{

\author{Burak Himmetoglu$^{(1)}$, Carlo R. Contaldi$^{(2)}$ and Marco Peloso$^{(1)}$}
\address{$^{(1)}${\it School of Physics and Astronomy, University of Minnesota, Minneapolis, MN 55455, USA}}
\address{$^{(2)}${\it Theoretical Physics, Blackett Laboratory, Imperial College, London, SW7 2BZ, UK}}

\begin{abstract}
We prove that the anisotropic inflationary background of the
Ackerman-Carroll-Wise model, characterized by a fixed-norm vector field, is unstable.  We found the instability by 
explicitly solving the linearized equations for the most general set of 
perturbations around this background, and by noticing that the solutions diverge close to
horizon crossing. This happens because one perturbation becomes 
a ghost at that moment. A simplified computation,  with 
only the perturbations of the vector field included, shows the same instability, clarifying the origin of the problem. We then discuss several other models, with a particular emphasis on the case of a 
nonminimal coupling to the  curvature, in which vector
fields are used either to support an anisotropic expansion, or to generate cosmological
perturbations on an isotropic background. In many cases, the mass term of the vector
needs to have the ``wrong'' sign; we show that, as a consequence, the longitudinal vector mode is 
a ghost (a field with negative kinetic term, and negative energy; not simply a tachyon). We comment
on problems that arise at the quantum level. In particular, the presence of a ghost
can be a serious difficulty for the UV completion that such models require in the sub-horizon regime.
\end{abstract}

\maketitle

\section{Introduction}
\label{sect-intro}

Cosmological observations over the past decade have ushered in the
long heralded era of precision cosmology with theoretical models of
the universe now being compared quantitatively to the data. In
particular, Cosmic Microwave Background (CMB) anisotropies
measurements from both orbital and sub-orbital experiments and Large
Scale Structure (LSS) observations have revealed a universe in overall
agreement with the standard inflationary paradigm. This picture
assumes an epoch preceding the standard, radiation dominated era when
the scale factor increased exponentially with a nearly constant Hubble
rate. This quasi deSitter expansion not only provides a solution to
the classical cosmological problems (homogeneity, isotropy, monopole
and flatness) but also results in a nearly scale-invariant spectrum of
super-horizon curvature perturbations which seed the structure
formation process.

However, certain features of the full sky CMB maps recently observed
by the WMAP satellite \cite{WMAP} seem to be in conflict with the
standard picture. These so called `anomalies' include the low power in
the quadrupole moment~\cite{cobe,wmap1,lowl}, the alignment of the
lowest multipoles, also known as the `axis-of-evil'~\cite{axis}, and
an asymmetry in power between the northern and southern ecliptic
hemispheres~\cite{asym}. The statistical significance of these effects
has been debated extensively in the literature. For instance, 
the significance of the anomalous lack
of large-angle correlations, together with the alignment of power has
also grown in strength with the latest data~\cite{Copi:2008hw}, with
only one in 4000 realizations of the concordance model in agreement
with the observations.

Recent studies on properly masked data have shown that an
anisotropic covariance matrix fits the WMAP low-$\ell$ data at the
3.8$\sigma$ level~\cite{Groeneboom:2008fz}. 
\footnote{An upper bound on the anisotropy has instead been obtained in
Ref.~\cite{ArmendarizPicon:2008yr}.} The fit was done using the
full anisotropic covariance matrix and found a best-fit anisotropy at
the $10\%$ level. The increase in significance arises from the fact
that much information is contained in the off-diagonal terms of the
covariance matrix in these models \cite{gcp,gcp2} and this is missed
by simple fits to the diagonal power spectrum.

Such violations of statistical isotropy are considered at odds with
the standard phase of early inflation. However, an
albeit more plausible explanation of these anomalies arising from a
systematic effect or foreground signal affecting the analysis is not
forthcoming.  This has led to a number of attempts at reconciling some
of the anomalies with the standard inflationary picture through
various modifications.

It has been suggested in~\cite{gcp,gcp2} that the alignment of lowest
multipoles could be could be due to an early anisotropic expansion.
Indeed, an anisotropic expansion results in a nonvanishing correlation
between different $a_{\ell m}$ coefficients of the CMB multipole
expansion (the formalism for dealing with the cosmological
perturbations on an anisotropic background was also developed
in~\cite{uzan1,uzan2}). To realize this with a minimal departure from
the standard inflationary picture, the anisotropy was simply set as an
initial condition at the onset of inflation. It was then shown in
\cite{gkp} that one of the gravity waves polarization experiences a
large growth during the anisotropic era (this is intimately related to
the instability of Kasner spaces \cite{gkp}), which may result in a
large $B$ signal in the CMB. Inflation however rapidly removes the
anisotropy.  The modes that leave the horizon well after the universe
has isotropized were deep inside the horizon while the universe was
anisotropic, and one recovers a standard power spectrum at the
corresponding scales. The signature of the earlier anisotropic stage
are therefore present only on relatively large scales, which were
comparable to the horizon when the universe was anisotropic. If the
duration of inflation was too long, such scales get inflated well
beyond the present horizon, and are irrelevant for
phenomenology. Therefore, the above suggestions require a tuned
duration of inflation.

To avoid this tuning, one can obtain prolonged anisotropic inflationary
solutions by introducing some ingredients that violate the premises of
Wald's theorem~\cite{wald} on the rapid isotropization of Bianchi
universes. This has been realized through the addition of quadratic
curvature invariants to the gravity action~\cite{barrow}, with the use
of the Kalb-Ramond axion~\cite{nemanja}, or of vector
fields~\cite{ford}. \footnote{The possibility of a late time
anisotropic expansion has instead
been considered in~\cite{dark}. Although the present statistics of the observed Supernovae does not show any evidence for the anisotropy~\cite{SN}, such studies are motivated by
the large increase of data that is expected in the next few years.} The basic idea underlying these
models is that one or more vector fields acquire
a spatial expectation value (vev). The background solutions are
homogeneous, so that the vev is only time dependent, and aligned along
one direction; for definiteness, we take it to coincide with the
$x-$axis. The models differ from the way in which the vector field(s)
acquires its vev. Most of this work we focus on one of the above models:
namely the model by Ackerman, Carroll, and Wise (ACW) \cite{acw}, in
which the vector acquires a fixed norm due to the constraint equation
enforced by a lagrange multiplier. There is also a cosmological
constant in the model, which is responsible for the accelerated
expansion. The vev of the vector field controls the difference between
the expansion rate of the $x$-direction and that of the two orthogonal
directions.

The original paper studied the power spectrum generated for a test
field $\delta \chi$ in this background. It was then argued that this
spectrum gets transferred to the spectrum of primordial density
perturbations through the mechanism of modulated perturbations
\cite{modulated}. The basic idea is as follows: imagine that the
effective cosmological constant in the model is replaced by an
inflaton field, and that the decay rate of the inflaton is controlled
by the vacuum expectation value of the test field $\chi$. Then,
fluctuations $\delta \chi$ will result in fluctuations of the decay
rate. This generates fluctuations of the energy density of the decay
products (since the energy density of the inflaton and of the decay
products scale differently with time). The resulting power spectrum
was the starting point of the analysis of
Ref.~\cite{Groeneboom:2008fz} cited above.  The claim
of~\cite{Groeneboom:2008fz} renders the ACW model particularly
interesting. However, the mechanism of modulated perturbations
requires that the perturbations of the metric and of the energy
densities which are unavoidably generated during inflation are
subdominant; otherwise the resulting density perturbations will be a
superposition of the original ones, generated without the test field,
plus those imprinted by $\delta \chi$ at reheating. The original ACW
work computed the perturbations of $\chi$ on an unperturbed
background. The underlying assumption is that the original
perturbations in the metric and in the energy densities can be
neglected with respect to $\delta \chi \,$. In this work we show that,
unfortunately, this assumption does not hold. On the contrary, each
mode (in Fourier space) of the linearized perturbations (which are
unavoidably generated irrespectively of the test field $\chi$) becomes
nonlinear while it leaves the horizon during inflation. As a
consequence, the background solution of \cite{acw} is unstable, and
any phenomenological study based on it is unreliable.

The computation of the instability is rather complicated, due to the
large number of perturbations involved. An important step in this
computation was performed in Ref.~\cite{dgw},~\footnote{Another interesting analysis can be found in Ref.~\cite{dg}, where the late time stability of the ACW model was studied, assuming that the cosmological constant is replaced by a fluid.} where the linearized
equations for the complete set of perturbations of the ACW model were
presented. These equations were solved in~\cite{dgw} in the
approximation of very large or very small wavelength of the
perturbations. Since the Hubble expansion rates for the ACW model are
constant, the wavelength of any mode increases exponentially with
time. Therefore, at sufficiently early times during inflation, the
wavelength is much smaller than the inverse hubble rates; this is the
small wavelength regime. On the contrary, at sufficiently late times
during inflation, the wavelength is much greater than the inverse
hubble rates; this is the large wavelength regime. The moment at which
those length become equal is known as horizon crossing~\footnote{The
  language is slightly loose, since there are two different expansion
  rates for the $x-$direction, and for the perpendicular $y-z$
  plane. However, we have in mind a small anisotropy (since a large
  one is forbidden by observations), so that the two rate are
  parametrically identical.}. We show here that the linearized
perturbations diverge close to horizon crossing, where the equations
were not solved in~\cite{dgw} (apart from this, our computation agrees
with that of~\cite{dgw}). This computation is presented in Section
\ref{sect-linearized}. It is just a conceptually straightforward
(although technically involved) computation and explicit solution of
the equations for the complete linearized system of perturbations for
the ACW model. Each mode of the perturbations becomes unstable at a
time $t_*$ close to horizon crossing (since different modes have
different wavelengths, they all become unstable at different
times). This analysis is all one needs to prove that the ACW
background solution is unstable.

To understand the reason for the instability, in Section~\ref{sect-ghost} we compute
the quadratic action for the perturbations. After the action is
diagonalized, we find that the kinetic term for one of the modes goes
from positive to negative precisely at the time $t_*$. This has two
consequences for the perturbations: 1) the equations of motion for the
perturbations diverge at this point, as we have seen; 2) the mode is a
ghost for $t>t_* \,$ (we note that this second point alone renders the
model unstable, due to the decay of the vacuum in ghost-nonghost pairs,
with a rate that increases with the momentum of the produced
particles). This last point is very interesting in the light of the
findings of Ref.~\cite{clayton}. It was shown there that, in models in
which the norm of the vector field is fixed as in the ACW model, it is
possible to construct field configurations with arbitrary large
negative energy. Those configurations were obtained ignoring
gravity. However, combining the result of Ref.~\cite{clayton} and
ours, we are inclined to conclude that such models are generally
unstable due to ghosts.

In Section~\ref{sect-simplified} we present a simplified computation, with only the perturbations of the vector field included (no metric perturbations). This is by itself an inconsistent computation, since metric perturbations are unavoidably sourced by the perturbations of the vector field. However, we show it for illustrative purposes, since it is much simpler than the complete computations presented in the first two parts. The qualitative results of this simplified computation are in perfect agreement with those of the complete one (the solutions diverge close to horizon crossing, when one mode becomes a ghost). This can be seen with extremely simple algebra: the mode that becomes a ghost is the longitudinal perturbation of the vector field (``longitudinal'' in the $y-z$ plane; 
see Section~\ref{sect-simplified} for the technical detail). Including the perturbations of the metric (as was done in the first two parts of this paper) simply ``dresses up'' the problem with technical complication, but does not remove the instability of the model, which is ultimately related to the vector field.

The instability found for the ACW model motivated us to study also different models in which vector fields with broken U(1) invariance play a role during inflation. In Section \ref{sect-other}, we specifically focus on models in which the mass term is due to a nonminimal coupling of the vector to the curvature
 \cite{Dimopoulos1,mukhvect,soda1}. These models also present a ghost. This is not related to an anisotropic expansion (in fact, the background is isotropic for \cite{Dimopoulos1}, in which the vector fields do not contribute to the expansion, and and can be made isotropic for \cite{mukhvect} by using $3N$ mutually orthogonal vectors with equal vev), but rather to the specific sign that the mass term needs to have in these models. This is due to the fact that, also in this case, the ill-behaved mode is the longitudinal polarization of the vector, which is present (and, therefore, acquires a kinetic term) only when the mass is nonvanishing. Since this is not a trivial point, as since this has not yet been pointed out for the above models, we present three different proofs of this statement. In addition to the ghost, we have verified that also for the models  \cite{mukhvect,soda1} the linearized equations for the perturbations diverge close to the horizon crossing, precisely as in the ACW model. The detailed computations will be presented elsewhere.
 
In Section \ref{sect-conclusion} we present some general discussions of these models. We do not provide quantitative computations here, as in the rest of the paper, but we present some comments and some worries that we believe deserve consideration. Particularly, we comment on some of the implications of having a ghost state in the theory. In the above models, the mass term provides a hard breaking of the gauge invariance. The interactions of the longitudinal vector polarization become strong at an energy scale parametrically set by the mass of the vector. This will occur for all the interactions involving the longitudinal vector, or mediated by it, renormalizing the values of the couplings between all the fields in the theory. Since the mass in these models is the hubble rate, or below, this may invalidate the predictions that rely on the initial quantization in the short wavelength regime (depending on the precise point in which the theory goes out of control). The most immediate solution to this problem is to provide a UV completion of the theory, in which the gauge symmetry is broken spontaneously from the condensation of a field that becomes dynamical above that mass scale. However, providing a UV completion appears to be a harder task when the longitudinal vector is a ghost rather than a normal field (for instance, in the higgs mechanism, the field responsible for the mass would need to be itself a ghost). We believe that such issues deserve further study.

\smallskip

Most of the computations presented here we summarized in our previous work \cite{hcp1}.

\section{Instability from the linearized equations}
\label{sect-linearized}

In this Section we write down and solve the linearized equations for the most general set of perturbations of the ACW background solution. The model, and the background solution, are presented in Subsection \ref{subsect-background}. In Subsection \ref{subsect-formalism} we introduce the perturbations; we show how to classify them in two distinct sets that are decoupled from each other at the linearized level, and how to construct gauge invariant variables. In Subsection \ref{subsect-linsol} we write down and solve the linearized equations for the gauge invariant combinations. Some intermediate steps of this computation are put in Appendix \ref{appexplicit}.

\subsection{The ACW model}
\label{subsect-background}

The ACW model is defined by the action,
\begin{equation}
S = \int d^4x\, \sqrt{-g} \left[ \frac{M_p^2}{2}\, R  -
\frac{1}{4} F_{\mu \nu} \, F^{\mu \nu} + \lambda \, \left( A^2 -
m^2 \right) - V_0 \right] \label{action-acw}
\end{equation}
where $A^2 \equiv A_{\mu}\,A^{\mu}$ and $V_0$ is a constant vacuum
energy (which is assumed to approximate a slow-roll inflationary
phase). This action is a special case of the one considered
in~\cite{acw}, which contains generalized kinetic terms.~\footnote{Specifically, the general
kinetic term for the vector field considered in~\cite{acw} is 
${\cal L} = - \beta_1 \nabla^\mu A^\nu \nabla_\mu A_\nu - 
\beta_2 \left( \nabla_\mu A^\mu \right)^2 - \beta_3 \, \nabla^\mu A^\nu \nabla_\nu A_\mu$. Ref. \cite{dgw} showed that the cases $\beta_1 < 0$ and $\beta_1 + \beta_2 + \beta_3 \neq 0$ have ghosts. We have restricted our attention to the case of a standard kinetic term, with $\beta_1 = - \beta_3 = 1/2 ,\, \beta_2 = 0 \,$.}

The evolution of the system is governed by
the field equations derived from the action (\ref{action-acw}):
\begin{eqnarray}
&& G_{\mu\nu} = \frac{1}{M_p^2}\, \left[ F_{\mu\alpha}\,
F_{\nu}^{\,\, \alpha} - 2 \lambda\, A_{\mu}\, A_{\nu} +
g_{\mu\nu}\, \left( -\frac{1}{4}\, F^2 + \lambda\, \left( A^2 -
m^2 \right) - V_0 \right) \right] \nonumber\\
&& \frac{1}{\sqrt{-g}}\, \partial_{\nu}\, \left[ \sqrt{-g}\,
F^{\mu\nu} \right] = 2\, \lambda\, A^{\mu} \nonumber\\
&& A^2 = m^2 \label{field-eqns-acw}
\end{eqnarray}
which are, respectively, the Einstein equations, the equation
of the vector field, and the constraint forced by the lagrange multiplier.
The constraint enforces a fixed norm of the vector.

The background metric and vector field are 
\begin{eqnarray}
\langle g_{\mu \nu} \rangle &=& {\rm diag } \left( -1 ,\, a \left( t \right)^2 ,\, b \left( t \right)^2 ,\, b \left( t \right)^2 \right) \nonumber\\
\langle A_\mu \rangle &=& \left( 0 ,\, M_p \, a \left( t \right) \, B_1 \left( t \right) ,\, 0 ,\, 0 \right) 
\label{background}
\end{eqnarray}
where in the vev for the vector field $\langle A_1 \rangle \equiv M_p \, a \, B_1$ we rescaled
out the scale factor $a$ and the reduced Planck mass $M_p$ for algebraic convenience (since the background equations of motion are simpler when written in terms of $B_1 \,$; we also note that $B_1$ is a dimensionless quantity). The last of (\ref{field-eqns-acw}) enforces a constant value for $B_1$,
\begin{equation}
B_1  = \frac{m}{M_p} \equiv \mu
\label{vector-back-acw}
\end{equation}

Denoting the two expansion rates $H_a \equiv \dot{a}/a$
and $H_b \equiv \dot{b}/b$, where dot is a time derivative,
the nontrivial equations in (\ref{field-eqns-acw}) are \footnote{Specifically, 
eqs. (\ref{eqns-conformal-acw}) are, respectively,
the $00 \,$, $11 \,$, $22$ Einstein equations, and the $x-$ component of the
vector field equation.}
\begin{eqnarray}
&& 2 H_a\, H_b + H_b^2 = \frac{1}{2}\, H_a^2\, \mu^2 +
\frac{V_0}{M_p^2} \nonumber\\
&& 2 \dot{H}_b + 3 H_b^2 = \frac{1}{2}\, H_a^2\, \mu^2 +
\frac{V_0}{M_p^2} + 2\, \lambda\, \mu^2 \nonumber\\
&& \dot{H_a} + \dot{H}_b + H_a^2 + H_a\, H_b + H_b^2 =
-\frac{1}{2}\, H_a^2\, \mu^2 + \frac{V_0}{M_p^2} \nonumber\\
&& \lambda = H_a\, H_b + \frac{1}{2}\, \dot{H}_a
\label{eqns-conformal-acw}
\end{eqnarray}
The equations are solved by exponentially expanding scale factors $a$ and $b$, with 
constant Hubble rates 
\begin{equation}
H_a^2 = \frac{2\, V_0}{M_p^2}\, \frac{1}{6 + 7 \mu^2 + 2 \mu^4}
\,\,\, , \,\,\, H_b = \left( 1 + \mu^2 \right)\, H_a
\label{Hab-acw}
\end{equation}
The rate $H_b$ which characterizes the expansion of the two
dimensional $y-z$ plane is larger than $H_a$, the expansion rate
of the anisotropic $x$-direction. The difference is proportional to
the background expectation value of the vector field. The overall inflationary
expansion is also supported by the vacuum energy $V_0 > 0 \,$.

\subsection{Classification of the perturbations and gauge invariant combinations}
\label{subsect-formalism}

We generalize to the present case the standard computation of cosmological perturbations in the case of a scalar field on an isotropic background (see \cite{mfb} for a review). It is convenient to exploit the symmetry of the background in this computation. For instance, in the isotropic case it is customary to 
classify the perturbations in scalar / vector / tensor modes, according to how they transform under spatial rotations. The underlying reason is that modes that transform differently are decoupled from each other at the linearized level (namely, in the linearized equations for the perturbations, or, equivalently, in the quadratic expansion of the action in the perturbations). Since the modes in that decomposition form a complete basis for the perturbations, we could still use them here. However, this would be of no practical advantage, since modes belonging to different representations would now be coupled to each other. We can however still exploit the residual symmetry of the ACW background solution (\ref{background}) in the $y-z$ plane, and classify the perturbations according to how they transform with respect to rotations in this plane \cite{gcp2}. Modes transforming differently under these rotations are decoupled at the linearized level (we explicitly verify this in the equations of Subsection~\ref{subsect-linsol}).

Specifically, we write the most general perturbations of the metric and of the vector field as
\begin{eqnarray}
\delta g_{\mu \nu} &=& 
 \left( \begin{array}{ccc}  -2 \Phi & a\,
\partial_1\, \chi & b\left( \partial_i\, B + B_i \right) \\
 &  - 2 \, a^2 \, \Psi  & a\, b\, \partial_1\, \left(
 \partial_i\, \tilde{B} + \tilde{B}_i \right) \\
  & & b^2\, \left(  - 2\Sigma \, \delta_{ij} - 2\,
  \partial_i\, \partial_j\, E - \partial_i\, E_i - \partial_j\,
  E_i \right) \end{array} \right) \nonumber\\ \nonumber\\
 \delta A_\mu &=& \left( \alpha_0 ,\, \alpha_1 ,\, \partial_i \alpha + \alpha_i \right)
\label{defn-perts-2d}
\end{eqnarray}
where $i,j = 1,2$ span the isotropic coordinates.~\footnote{We chose to insert a $\partial_1$ derivative in the $\delta g_{01}$ and $\delta g_{1i}$ metric perturbations for algebraic convenience. For instance, we could have equivalently denoted $\delta g_{01} = \chi_1$. More precisely, we only study the modes of the perturbations which have a nonvanishing momentum component both along the $x$ direction and in the  $y-z$ plane (in coordinate space, they have a nontrivial dependence both on $x$ and at least on one between $y$ and $z$).} The perturbations $\{ \Phi, \, \chi, \, B, \, \Psi, \, \tilde{B}, \, \Sigma, \, E, \, \alpha_0, \, \alpha_1, \, \alpha \}$ are $2$d scalar modes: they encode $1$ degree of freedom (d.o.f.) each. The perturbations $\{ B_i,\, \tilde{B}_i, \, E_i,\, \alpha_i \}$ are $2$d vector modes. Due to the transversality condition ($ \partial_i B_i = \dots = 0 $), they also encode $1$ d.o.f. each. Notice that, contrary to the $3$d case, there are no $2$d tensor modes, since the transversality and traceless conditions eliminate all possible degrees of freedom. Altogether, we have $10$ d.o.f. in the $2$d scalar sector, and $4$ d.o.f. in the $2$d vector sector. These add up to $14$ d.o.f., which is the number of initial independent entries in the metric and in the vector field.

The set of perturbations just given is redundant, since modes can be transformed into each other
through infinitesimal coordinate transformations $x^\mu \rightarrow x^\mu + \xi^\mu$. Under this change
\cite{mfb}
\begin{equation}
\delta g_{\mu \nu} \rightarrow \delta g_{\mu \nu} - \langle g_{\mu \nu ,\alpha} \rangle \xi^\alpha -
\langle g_{\mu \alpha} \rangle \xi^\alpha_{,\nu} - \langle g_{\alpha \nu} \rangle \xi^\alpha_{,\mu} \;\;\;,\;\;\;
\delta A_\mu \rightarrow \delta A_\mu - \langle A_{\mu ,\alpha} \rangle \xi^\alpha - \langle A_\alpha \rangle \xi^\alpha_{,\mu}
\label{gaugetransf}
\end{equation}
We decompose also the components of the infinitesimal parameter $\xi^\mu$ in the $y-z$ plane in a $2$d scalar plus $2$d vector part. In this way $2$d scalar (vector) modes manifestly transform into $2$d scalar (vector) modes.

To eliminate the redundancy, we can specify a gauge that completely fixes the freedom of general coordinate reparametrization (this was the procedure chosen in \cite{gcp2}, where the choice $E = \Sigma = {\tilde B} = E_i = 0$ was made). Equivalently, we can construct gauge invariant combinations of the above perturbations (``gauge invariance'' here means invariance with respect to general coordinate transformations; there is no gauge U(1) symmetry associated to the vector field, due to its mass term). 
We choose this second procedure here: each linearized equation can be written as ``left hand side $= 0$''; therefore the equation must be gauge invariant. This means that the perturbations must enter in  the linearized equations in such a way that these equations can be written in terms of the gauge invariant combinations only (the same is true for the quadratic action that we compute in the next Section). This provides a nontrivial check on our algebra.

To find the gauge invariant combinations, we compute how each mode in
(\ref{defn-perts-2d}) transforms under (\ref{gaugetransf}), and we
then arrange them into invariant combinations. We find the following
set of gauge invariant modes 
\begin{eqnarray}
\hat{\Phi} &=& M_p\, \left[ \Phi + \left( \frac{\Sigma}{H_b}
\right)^{\bullet} \right] \nonumber\\
\hat{\Psi} &=& M_p\, \left[ \Psi - \frac{H_a}{H_b}\, \Sigma +
\frac{b}{a}\, \partial_1^2\, \left( \tilde{B} + \frac{b}{a}\, E
\right) \right] \nonumber\\
\hat{B} &=& -\frac{M_p^2}{b}\, \vec{\partial}_T^2\, \left[ B -
\frac{1}{b\, H_b}\, \Sigma + b\, \dot{E} \right] \nonumber\\
\hat{\chi} &=& -\frac{M_p}{a}\, \partial_1^2\, \left[ \chi -
\frac{1}{a\, H_b}\, \Sigma - a\, \left( \frac{b}{a}\, \left(
\tilde{B}+\frac{b}{a}\, E \right) \right)^{\bullet} \right]
\nonumber\\
\hat{\alpha}_1 &=& -\frac{1}{a}\, \left[ \alpha_1 + a\, M_p\,
\frac{\dot{B}_1 + H_a\, B_1}{H_b}\, \Sigma - b\, M_p\, B_1\,
\partial_1^2\, \left( \tilde{B} + \frac{b}{a}\, E \right) \right]
\nonumber\\
\alpha &=& -\frac{1}{a}\, \partial_1\, \left[ \alpha - b\, M_p\,
\partial_1\, \left( \tilde{B} + \frac{b}{a}\, E \right) \right]
\nonumber\\
\hat{\alpha}_0 &=& \frac{1}{a}\, \partial_1\, \left[ \alpha_0 -
a\, M_p\, B_1\, \partial_1\, \left( \frac{b}{a}\, \left( \tilde{B}
+ \frac{b}{a}\, E \right) \right)^{\bullet} \right] \label{GI-2dS}
\end{eqnarray}
(where the bullet denotes time derivative, and $\vec{\partial}_T^2 \equiv \partial_2^2 + \partial_3^2$) in the $2$d scalar sector, and 
\begin{eqnarray}
\hat{B}_i &=& B_i + b\, \dot{E}_i \nonumber\\
\hat{\tilde{B}}_i &=& a\left( \tilde{B}_i + \frac{b}{a}\, E_i
\right) \nonumber\\
\hat{\alpha}_i &=& \frac{\alpha_i}{M_p} \label{GI-2dV}
\end{eqnarray}
in the $2$d vector  sector. The prefactors in front of these modes do not affect their gauge invariance, but have been chosen for algebraic convenience.

Since we started from ten $2$d scalars and four $2$d vectors, and since there are three infinitesimal $2$d scalar transformations, and one infinitesimal $2$d vector transformation, we end up with seven $2$d scalar gauge invariant modes, and three $2$d vector gauge invariant modes. One could have equivalently chosen other gauge invariant combinations of the initial modes. However, they can be obtained from those given here.~\footnote{One can verify that the present choice of variables leads to the same results as the standard computation of \cite{mfb}, in the case of isotropic background, and of a scalar field instead of the vector. To do so, consider the gauge specified in \cite{gcp2} (in practice, it amounts in identifying each ``hatted'' mode with the corresponding ``non-hatted'' one). It was explicitly verified in \cite{gcp2} that, in the isotropic limit, this gauge leads to the same results as the longitudinal gauge. The perturbations in the longitudinal gauge can be immediately ``promoted'' to the gauge invariant modes used in \cite{mfb}. This explicitly verifies that our formalism reproduces the standard one in the isotropic limit.} 

\bigskip

We perform the computations in momentum space:
\begin{equation}
\delta \left( x \right) = \frac{1}{\left( 2 \pi \right)^3} \int d^3 k \, \delta \left( k \right) {\rm e}^{-i \vec{k} \, \vec{x}}
\label{Fourier}
\end{equation}
where $\delta$ denotes any of the above perturbations, and where we denote with the same symbol the perturbation both in coordinate and in momentum space. $k$ is the comoving momentum of the mode $\delta \left( k \right)$. Different modes (characterized each by a specific comoving momentum) are decoupled from each other at the linearized level. We denote by $k_L$ the component of the comoving momentum in the $x-$direction, and by $k_T$ the component in the orthogonal $y-z$ plane. The corresponding components of the physical momentum are $p_L = k_L / a \left( t \right)$ and $p_T = k_T / b \left( t \right)$. The square magnitude of the comoving/physical momenta are given by $k^2 = k_L^2 + k_T^2$ and $p^2 = p_L^2 + p_T^2 \,$, respectively. Finally $k_{Ti}$ ($p_{Ti}$) denotes the component of the transverse comoving (physical) momentum in the $y$ ($i=2$) or $z$ ($i=3$) direction.

In momentum space, the transversality conditions on the $2$d vectors reads
\begin{equation}
k_{Ti} v_i = p_{Ti} v_i = 0
\end{equation}
where $v_i$ denotes any of the $2$d vector perturbations.

\subsection{Solutions to the linearized equations for the perturbations}
\label{subsect-linsol}

We perturb the metric and the vector field as in eq. (\ref{defn-perts-2d}). We want to solve the system of linearized equations for these modes. We multiply the second line of (\ref{field-eqns-acw}) by $A_\mu$ and we replace $A^2$ by $m^2$ on the right hand side. In this way we obtain an expression for 
$\lambda$ in terms of the vector field and the metric (this is the procedure also adopted in 
\cite{Carroll:2004ai} to study a similar model, and it has the advantage that we do not have to introduce explicitly the perturbation of the lagrange multiplier). In this way, the first and second line 
of (\ref{field-eqns-acw}) become
\begin{eqnarray}
&&G_{\mu\nu} = \frac{1}{M_p^2}\, T_{\mu\nu} = \frac{1}{M_p^2}\,
\left[ F_{\mu\alpha}\, F_{\nu}^{\,\, \alpha} - \frac{1}{\mu^2\,
M_p^2}\, A_{\alpha}\, \nabla_{\beta}\, F^{\alpha\beta}\, A_{\mu}\,
A_{\nu} + g_{\mu\nu}\, \left( -\frac{1}{4}\, F^2 - V_0 \right)
\right]  \nonumber\\
&&\partial_\nu \left[ \sqrt{-g} \, F^{\lambda \nu} \right] \left( m^2 \, \delta_\lambda^\mu - A_\lambda \, A^\mu \right) = 0
\label{eqs-acw-nol}
\end{eqnarray}
We then linearize these equations, and we go to momentum space as indicated in (\ref{Fourier}). 
We denote the resulting equations as \footnote{After we eliminated $\lambda$, the linearization of the third of (\ref{field-eqns-acw}) coincides with ${\rm Eq}_1$.}
\begin{equation}
{\rm Eq}_{\mu\nu}: \delta\, \left( G_{\mu\nu} - \frac{1}{M_p^2}\,
T_{\mu\nu} \right) = 0 \,\,\,\, , \,\,\,\,\, {\rm Eq}_{\mu}:
\delta\, \left( \frac{\delta S}{\delta A_{\mu}} \right) = 0
\label{pert-eqns-acw} 
\end{equation}
The explicit expressions are given in eqs. (\ref{pert-eqns-acw-exp}) in Appendix \ref{appexplicit}.
The expressions pass two crucial tests: we initially wrote them in terms of the original perturbations
(\ref{defn-perts-2d}); with some algebra, we have been able to write them solely in terms of the gauge invariant combinations (\ref{GI-2dS}) and (\ref{GI-2dV}). As a second test, we show explicitly in Appendix \ref{appexplicit} that the set of $2$d scalar and $2$d vector perturbations decouple in these equations. We verified that the system of $2$d vectors is stable. For brevity, we do not report those computations here. We also show in Appendix \ref{appexplicit} that not all the equations (\ref{pert-eqns-acw}) are independent (due to the perturbed Bianchi identities). A set of independent equations is
\begin{eqnarray}
{\rm Eq}_1 : && {\hat \alpha_1} - \mu \, {\hat \Psi} = 0
\nonumber\\ \label{Eq1} \\
{\rm Eq}_{00} : && \left( 2 + \mu^2 \right) H_a \, \dot{\hat\Psi}
+ p_T^2\, \hat\Psi + \left( 2 + \mu^2 \right)\, \left(3 +
2\mu^2\right)\, H_a^2 \, {\hat \Phi} \nonumber\\
&& \qquad\qquad\qquad - 2 \left( 1+\mu^2 \right) H_a \, {\hat
\chi} - \left( 2 + \mu^2 \right) H_a \, {\hat B} - \mu \, H_a \,
{\hat \alpha}_0 = 0 \label{Eq00} \\
{\rm Eq}_{01} : && 2 \left( 1 + \mu^2 \right) H_a {\hat \Phi} +
\frac{1}{2 \, p_L^2} \left[ p_T^2 - 4 \, \mu^2 \left( 1 + \mu^2
\right) H_a^2 \right] {\hat \chi} - \frac{\hat B}{2} - 2 \mu
\left( 1 + \mu^2 \right) \frac{H_a^2}{p_L^2} {\hat \alpha}_0 = 0
\label{Eq01} \\
{\rm Eq}_{0i} : && \dot{\hat\Psi} + \left( 2 + \mu^2 \right) H_a
\, \hat\Phi - \frac{\hat \chi}{2} + \frac{p_L^2}{2 p_T^2}\, \hat B
- \mu \, H_a \, {\hat \alpha}  = 0
\label{Eq02} \\
{\rm Eq}_{11} : && \mu^2 \, \ddot{\hat \Psi} + \mu^2 \left( 3 + 2
\, \mu^2 \right) H_a \dot{\hat \Psi} + \mu^2 \, p_T^2 \, {\hat
\Psi} - \left( 2 + \mu^2 \right) H_a \, \dot{\hat \Phi} + \left[
p_T^2 - \left( 2 + \mu^2 \right)\, \left(3 +
2\mu^2\right)\, H_a^2 \right] {\hat \Phi} \nonumber\\
&& + \dot{\hat B} + \left( 3 + 2 \, \mu^2 \right) H_a \, {\hat B}
+ \mu \, \dot{\hat \alpha}_0 + \mu \left( 3 + 2 \, \mu^2 \right)
H_a \, {\hat \alpha}_0 - \mu \, p_T^2 \, {\hat \alpha} = 0
\label{Eq11} \\
{\rm Eq}_0 : && \dot{\hat\alpha} + H_a\, \hat\alpha +
\frac{1}{p_T^2} \left[ p^2 - 2 \left( 1 + \mu^2 \right) H_a^2
\right] {\hat \alpha}_0 + \mu \, \frac{p_L^2}{p_T^2}\,
\dot{\hat\Psi} + \mu \, \frac{p_L^2}{p_T^2}\, H_a\, {\hat \Phi} -
\mu \left( 1 + \mu^2 \right) \frac{2 \, H_a^2}{p_T^2} \, {\hat
\chi}  = 0 \nonumber\\ \label{Eq0} \\
{\rm Eq}_i : && \ddot{\hat\alpha} + 3 H_a\, \dot{\hat\alpha} +
\dot{\hat\alpha}_0 + \left( p_L^2 - 2 \, \mu^2 \, H_a^2 \right)\,
\hat\alpha - \mu \, p_L^2\, \hat\Psi + \mu \,
\frac{p_L^2}{p_T^2}\, H_a\, \hat{B} + 2 H_a\, \hat{\alpha}_0 = 0
\label{Eq2}
\end{eqnarray}
where we remind that $p_L$ and $p_T$ denote, respectively, the component of the physical 
momentum of the perturbations along the $x$ direction and in the orthogonal $y-z$ plane. We 
have used (\ref{Eq1}) to eliminate ${\hat \alpha}_1$ in favor of ${\hat \Psi}$ in all the other equations. Therefore, we do not need to consider this equation further in the following.

\bigskip

We now solve the set of equations (\ref{Eq00})-(\ref{Eq2}). We need to rearrange them in a system that can be numerically integrated. There are several equivalent ways to proceed. We could either integrate all the above equations numerically, or we could solve some of them analytically. We choose this second option (in this way, we reduce the number of equations to be numerically integrated, but we generally obtain more involved expressions). We start by solving equations (\ref{Eq01}) and (\ref{Eq02}) for $\hat\chi$ and $\hat{B}$. First, we need to verify that this can be done. We see that this is the case, since
\begin{equation}
\Big\vert \frac{\partial \left\{ {\rm Eq}_{01} ,\, {\rm Eq}_{0i} \right\}}{\partial \left\{ {\hat \chi} ,\, {\hat B} \right\}} \Big\vert =- \frac{\mu^2 \left( 1 + \mu^2 \right) H_a^2}{p_T^2} \neq 0
\end{equation}
The explicit solutions are 
\begin{eqnarray}
{\hat \chi} &=& \frac{p_T^2}{2 \, \mu^2 \, \left( 1 + \mu^2
\right) H_a^2} \dot{\hat \Psi} + \frac{1}{\mu^2 \, H_a} \left[ p^2
- \frac{\mu^2}{2 \left( 1 + \mu^2 \right)} p_T^2 \right] {\hat
\Phi}- \frac{p_T^2}{2 \, \mu \, \left( 1 + \mu^2 \right) H_a} {\hat \alpha} - \frac{{\hat \alpha}_0}{\mu} \nonumber\\
{\hat B} &=& \frac{1}{2 \, \mu^2 \, H_a^2} \, \frac{p_T^2}{p_L^2}
\left( \frac{p_T^2}{1+\mu^2} - 4 \, \mu^2 \, H_a^2 \right)
\dot{\hat \Psi} + \frac{1}{\mu^2 \, H_a} \, \frac{p_T^2}{p_L^2}
\left[ p^2 - \frac{\mu^2}{2 \left( 1 + \mu^2 \right)} p_T^2 - 2
\left( 2 + \mu^2 \right) \mu^2 \,  H_a^2 \right]
{\hat \Phi} \nonumber\\
&&\quad\quad\quad-\frac{1}{2 \, \mu \, H_a} \, \frac{p_T^2}{p_L^2}
\left( \frac{p_T^2}{1+\mu^2} - 4 \, \mu^2 \, H_a^2 \right) {\hat
\alpha} - \frac{p_T^2}{p_L^2} \, \frac{{\hat \alpha}_0}{\mu}
\label{solBhch}
\end{eqnarray}

Inserting these solutions into the remaining equations (\ref{Eq00}),  (\ref{Eq11}),  
(\ref{Eq0}), and (\ref{Eq2}), we have now a system of $4$ equations in terms of the four unknown modes ${\hat \Psi} ,\, {\hat \alpha} ,\, {\hat \Phi}$, and ${\hat \alpha}_0 \,$. These equations are explicitly given in
appendix \ref{appexplicit}, eqs. (\ref{Eq00-2}) - (\ref{Eq2-2}). As we show in appendix \ref{appexplicit}, by simple algebraic manipulation of these equation we obtain the system
\begin{equation}
{\cal M}_\kappa \, \left( \begin{array}{c}
\ddot{\hat \alpha} \\  \dot{\hat \alpha}_0 \\ \ddot{\hat \Psi} \\
\dot{\hat \Phi}
\end{array} \right) = \left( \begin{array}{c}
f_1 \\ f_2 \\ f_3 \\ f_4 \end{array} \right) 
\;\;\;,\;\;\; {\cal M}_\kappa \equiv 
\left( \begin{array}{cccc}
1 & 1 & 0 & 0 \\
0 & \kappa_{22} & \kappa_{23} & \kappa_{24} \\
1 & \kappa_{32} & \kappa_{33} & \kappa_{34} \\
0 & \kappa_{42} & \kappa_{43} & \kappa_{44}
\end{array} \right)
\label{matrix}
\end{equation}
(for reasons explained in the next Section, there are no second time derivatives of ${\hat \alpha}_0$ and ${\hat \Phi}$ in these equations).  The entries of ${\cal M}_\kappa$, given in eq. (\ref{coeffk}) depend on background quantities, while the four expressions  at right hand side, given in eq. (\ref{coefff}), are linear combinations of the unknown quantities $\left\{ {\hat \alpha} ,\, \dot{\hat \alpha} ,\, {\hat \alpha}_0 ,\, {\hat \Psi} ,\, \dot{\hat \Psi} ,\, {\hat \Phi} \right\} $ (also the coefficients of these linear combinations depend on background quantities). It is now straightforward to invert ${\cal M}_\kappa$, and to integrate the system numerically.

It is easy to see that, in general, the solutions of (\ref{matrix}) diverge close to horizon crossing. Indeed, 
\begin{equation}
{\rm det } \, {\cal M}_\kappa = \frac{p_T^2}{p_L^2} \, \frac{1+\mu^2}{\mu^2} \, \left[ 
p_L^2-\left(2+\mu^2\right)\, H_a \, H_b \right]
\end{equation}
so that, when we invert this matrix, we encounter a singularity when $p_L = \sqrt{2+\mu^2} \, \sqrt{H_a \, H_b} \,$.

We evolve the system, starting from the adiabatic vacuum initial conditions deeply inside the horizon (eqs. (\ref{in-nondyn}) and (\ref{initialcond})). The two components of the physical momentum evolve as
\begin{eqnarray}
p_L \left( t \right) &=& \frac{k_L}{a \left( t \right)} = p_{L0} \, {\rm e}^{-H_a \, t} \nonumber\\
p_T \left( t \right) &=& \frac{k_T}{b \left( t \right)} = p_{T0} \, {\rm e}^{-H_b \, t} =
p_{T0} \, {\rm e}^{- \left( 1 + \mu^2 \right) H_a \, t}
\end{eqnarray}
where $p_{L0} ,\, p_{T0}$ are the values of these components when $t =0$, and where we have used the background relation (\ref{Hab-acw}) between the two Hubble constants. We are free to choose the origin of the time. We set $t=0$ at the moment in which ${\rm det } \, {\cal M}_\kappa = 0 \,$. This fixes $p_{L0} = \sqrt{2+\mu^2} \, \sqrt{H_a \, H_b} \,$. The mode is then completely specified by giving the ratio $p_{T0} / p_{L0} \,$. For definiteness, we choose $\mu = 0.1 \,$ (giving an ${\rm O} 
\left( 10^{-2} \right)$ anisotropy), and $p_{T0} = H_a$ (so that the two components of the momentum are comparable to each other in the time range considered). We start the numerical evolution at $t=-7 / H_a$, so that $H/p={\rm O}\left( 10^{-3} \right)$ initially, and the modes are well inside the horizon.

\begin{figure}[h]
\centerline{
\includegraphics[width=0.4\textwidth,angle=-90]{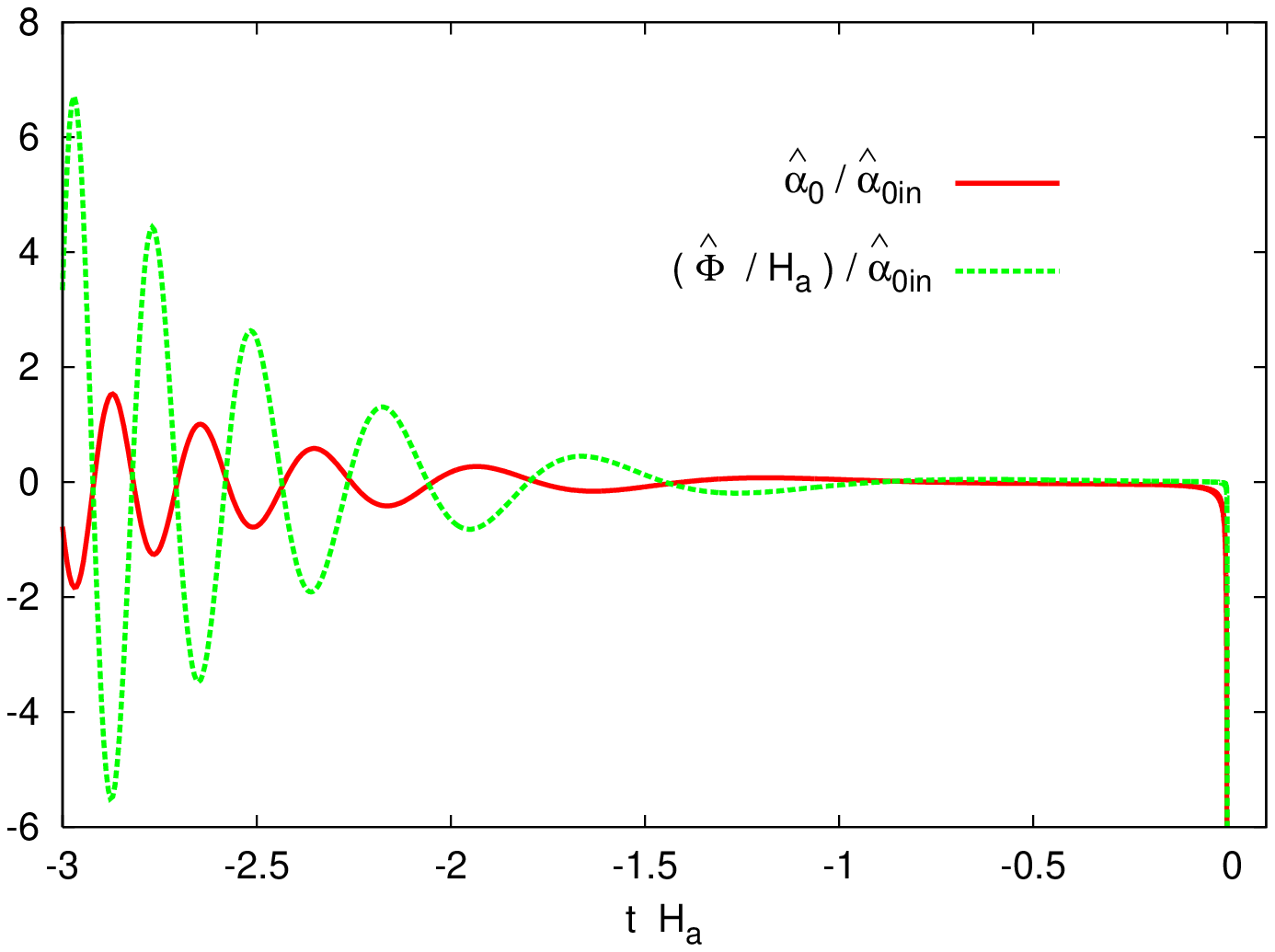}
\includegraphics[width=0.4\textwidth,angle=-90]{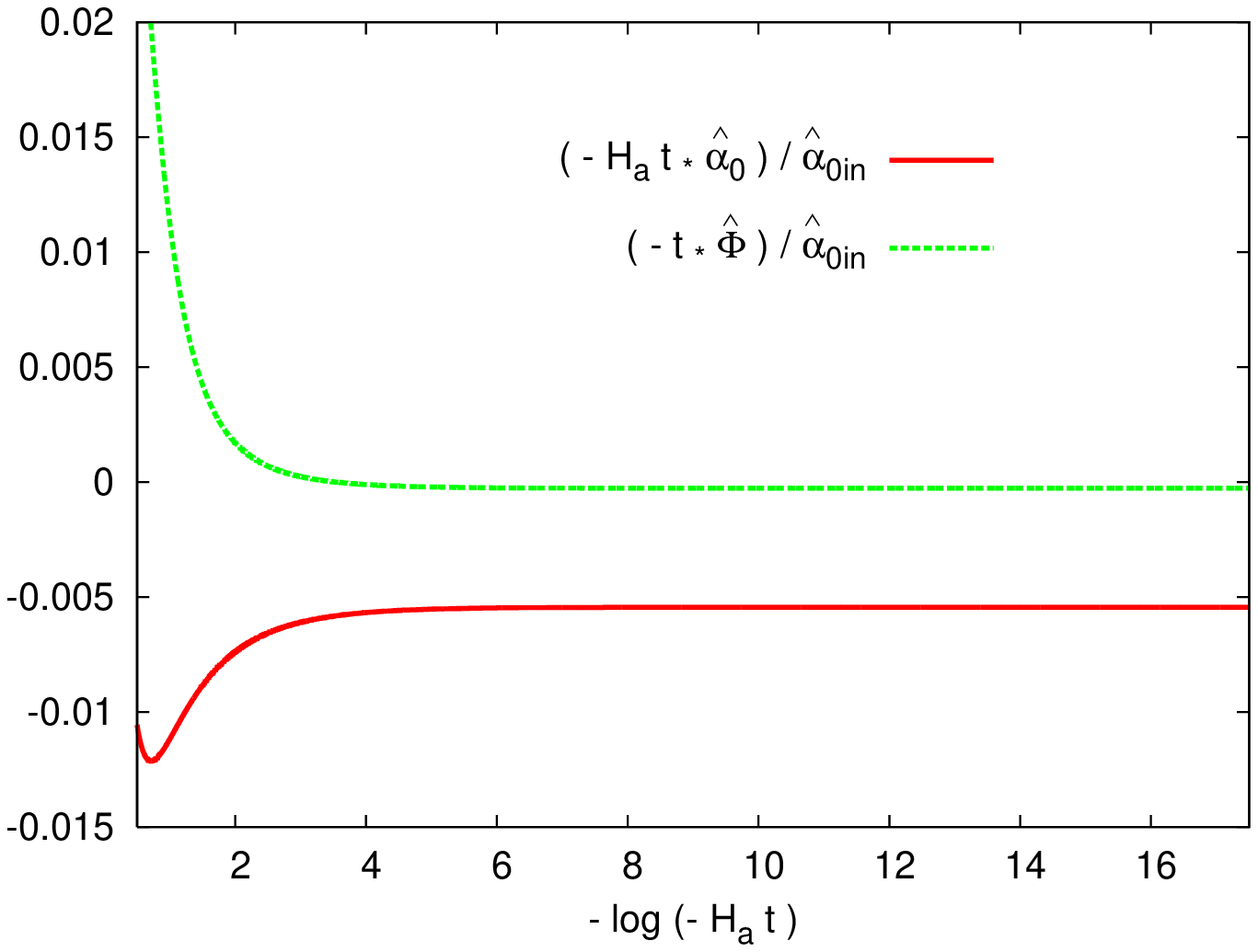}
} \caption{The right panel shows a zoom of the left panel in the
region $-0.6 \leq H_a \, t \leq - 10^{-5 } \,$ (the time is shown
in log units). Both modes show a 1/t divergence.} \label{fig:fig1}
\end{figure}

The left panel of Figure \ref{fig:fig1} shows that the modes ${\hat \alpha}_0$ and ${\hat \Phi}$ indeed diverge when ${\rm det } \, {\cal M}_\kappa = 0 \,$. The right panel gives the behavior close to the singularity. We give time in log units, so that $t=0$ is mapped at $- {\rm log } \left( - H_a \, t \right) = \infty \,$. We see that both ${\hat \alpha}_0$ and ${\hat \Phi}$ diverge linearly. The two modes
${\hat \alpha}$ and ${\hat \Psi}$ (not shown here) diverge logarithmically. The modes ${\hat \alpha}$ and ${\hat \alpha}_0$ present an identical divergence in the simplified computation of Section~\ref{sect-simplified}. For that simplified analysis, we have also obtained the divergence analytically. It is also clear from there that the solutions diverge for generic initial conditions (we have also verified this in the numerical solutions of the system (\ref{matrix})).

\section{Ghost from the quadratic equation}
\label{sect-ghost}

To gain a better understanding on the nature of the instability, we computed the quadratic action for
the perturbations. We inserted the expression (\ref{background}) and (\ref{defn-perts-2d}) in the action (\ref{action-acw}). We expanded at the second order in the perturbations. We disregarded the $2$d vector modes, which are decoupled from the $2$d scalars. We also eliminated the mode $\hat\alpha_1$ through the constraint equation (\ref{Eq1}). We then Fourier transformed the perturbations as in 
(\ref{Fourier}). The final result is, up to boundary terms,
\begin{eqnarray}
S_{{\rm 2dS}} &=& \frac{1}{2}\, \int d^3 k\, dt\, a\, b^2\, {\cal
L}_{{\rm 2dS}} \nonumber\\
{\cal L}_{{\rm 2dS}} &=& \mu^2\, \vert \dot{\hat\Psi} \vert^2 +
\frac{p_T^2}{p_L^2}\, \vert \dot{\hat\alpha} \vert^2 - H_a\,
\left( \hat{\Psi}^*\, \dot{\hat\Psi} + {\rm h.c.} \right) - \left(
2 + \mu^2 \right)\, H_a\, \left( \hat{\Phi}^*\, \dot{\hat\Psi} +
{\rm h.c.} \right) + \left( \hat{B}^*\, \dot{\hat{\Psi}} + {\rm
h.c.} \right) \nonumber\\
&& + \mu\, \left( \hat{\alpha}_0^*\, \dot{\hat\Psi} + {\rm h.c.}
\right) + \frac{p_T^2}{p_L^2}\, H_a\, \left( \hat{\alpha}^*\,
\dot{\hat\alpha} + {\rm h.c.} \right) + \frac{p_T^2}{p_L^2}\,
\left( \hat{\alpha}_0\, \dot{\hat\alpha} + {\rm h.c.} \right)
-\left[ \mu^2\, p_T^2 + \left( 3 + 2\mu^2 \right)\, H_a^2
\right]\, \vert \hat\Psi \vert^2 \nonumber\\
&& - p_T^2\, \left( \hat{\Phi}^*\, \hat\Psi + {\rm h.c.} \right) +
\mu\, p_T^2\, \left( \hat{\alpha}^*\, \hat\Psi +{\rm h.c.} \right)
- \left(2 + \mu^2 \right)\, \left( 3 +2 \mu^2 \right)\, H_a^2\,
\vert \hat\Phi \vert^2 \nonumber\\
&& + 2\, \left( 1 + \mu^2 \right)\, H_a\, \left( \hat{\chi}^*\,
\hat\Phi + {\rm h.c.} \right) + \left( 2 + \mu^2 \right)\, H_a\,
\left( \hat{B}^*\, \hat{\Phi} + {\rm h.c.} \right) + \mu\, H_a\,
\left( \hat{\alpha}_0^*\, \hat\Phi + {\rm h.c.} \right) \nonumber\\
&& + \frac{p_T^2 - 4 \mu^2\, \left( 1 + \mu^2 \right)\, H_a^2}{2
p_L^2}\, \vert \hat\chi \vert^2 - \frac{1}{2}\, \left( \hat{B}^*\,
\hat\chi + {\rm h.c.} \right) - \frac{2 \mu\, \left( 1 + \mu^2
\right)\, H_a^2}{p_L^2}\, \left( \hat{\chi}^*\, \hat{\alpha}_0 +
{\rm h.c.} \right) + \frac{p_L^2}{2 p_T^2}\, \vert \hat{B} \vert^2
\nonumber\\
&& - \mu\, H_a\, \left( \hat{B}^*\, \hat\alpha + {\rm h.c.}
\right) - \frac{p_T^2}{p_L^2}\, \left[ p_L^2 - \left( 3 + 2 \mu^2
\right)\, H_a^2 \right]\, \vert \hat\alpha \vert^2 +
\frac{p_T^2}{p_L^2}\, H_a\, \left( \hat{\alpha}_0^*\, \hat\alpha +
{\rm h.c.} \right) \nonumber\\
&& + \frac{p^2 - 2\, \left( 1 + \mu^2 \right)\, H_a^2}{p_L^2}\,
\vert \hat{\alpha}_0 \vert^2 \label{action-2dS-acw}
\end{eqnarray}
The computation is conceptually straightforward, although technically 
involved. The resulting action passes several nontrivial checks. Firstly, the
perturbations rearrange so that the action can be written solely in terms of gauge
invariant combinations. Secondly, extremizing the action with respect to these modes gives precisely the linearized equations (\ref{Eq00}) - (\ref{Eq2}). Thirdly, the action correctly identifies the nondynamical variables of the system. The nondynamical modes are the modes entering in the action (\ref{action-2dS-acw}) without time derivatives. They do not correspond to propagating dynamical degrees of freedom (since the nondynamical modes can be obtained from the dynamical ones by their own equations of motion, that are algebraic - since they enter without time derivatives in the action).

Whether a mode is dynamical or not depends on the initial kinetic
terms. For instance, from the field strength $F_{\mu \nu}
= \partial_\mu A_\nu - \partial_\nu A_\mu$ we immediately see that the
temporal component of the vector field is nondynamical. By direct
inspection, we can also verify that the Ricci scalar does not lead to
time derivatives for any of the $g_{0\mu}$ entry of the metric
(this is most easily seen through the ADM formalism \cite{adm}). The
gauge choice of \cite{gcp2} ($E = \Sigma = {\tilde B} = E_i = 0$ ) preserves 
all the $\delta g_{0\mu}$ modes, as well as $\delta
A_0 \,$, namely the modes $\Phi ,\, \chi ,\, B ,\, B_i ,\, \alpha_0$. 
These perturbations are the nondynamical fields in that
gauge. Therefore, the combinations ${\hat \Phi} ,\, {\hat \chi} ,\,
{\hat B} ,\, {\hat B}_i ,\, {\hat \alpha}_0$ given in (\ref{GI-2dS})
and (\ref{GI-2dV}) are nondynamical (since they are gauge invariant,
and since they reduce to the nondynamical modes of the system once the
gauge choice of \cite{gcp2} is made). This is explicitly verified by
the quadratic action (\ref{action-2dS-acw}). We integrate the
nondynamical modes out of the action (\ref{action-2dS-acw}). Namely,
we express them as a function of the dynamical modes by using their
own algebraic equations of motion. It is instructive to compare this
with what happens in the standard case of a scalar field in an
isotropic background. In that case, one starts from $11$ perturbations
(ten in the metric, and one in the scalar field). Four of them are
removed by the gauge fixing. Four of them are nondynamical, and are
integrated out (for instance, this is what is done in Ref.~\cite{mfb}
when the constraint equation (10.53) is used in going from (10.57) to
(10.59)). One ends up with three physical propagating modes: the two
polarizations of the tensor mode (gravity waves) and the scalar
density contrast. In general it is not immediately obvious which
combination of gauge invariant perturbations are dynamical (for
instance, this is not manifest in the choice of \cite{mfb}). However,
our choice of ``hatted'' modes makes this manifest.

To be more explicit, the $2$d modes entering in (\ref{action-2dS-acw})
can be divided in the two sets of dynamical $Y= \left\{ {\hat \Psi}
  ,\, {\hat \alpha} \right\}$ and nondynamical $N = \left\{ {\hat
    \Phi} ,\, {\hat \chi} ,\, {\hat B} ,\, {\hat \alpha_0} \right\}$
modes. The action (\ref{action-2dS-acw}) is formally of the type
\begin{equation}
S_{YN}^{(2)} = \int d t \, d^3 k \left[ a_{ij} \, \dot{Y}_i^* \, \dot{Y}_j + \left( b_{ij} \, N_i^* \, \dot{Y}_j + {\rm h. \, c.} \right) + c_{ij} \, N_i^* \,  N_j + \left( d_{ij} \, \dot{Y}_i^* \, Y_j + {\rm h. \, c.} \right) + e_{ij} \, Y_i^* \, Y_j + 
\left( f_{ij} \, N_i^* \, Y_j  + {\rm h. \, c.} \right) \right]
\label{genYNact}
\end{equation}
where the matrices formed by the $a,c,e$ coefficients are hermitian. The equations of motion for the nondynamical modes are
\begin{equation}
\frac{\delta S_{YN}^{(2)}}{\delta N_i^*} = 0 \;\;\; \Rightarrow \;\;\;
c_{ij} \, N_j = - b_{ij} \, \dot{Y}_j - f_{ij} \, Y_j
\label{intout}
\end{equation}
These equations are precisely Eqs. (\ref{Eq00}), (\ref{Eq01}), (\ref{Eq02}), and (\ref{Eq0}) of the linearized system that we have solved (notice that, indeed, they not contain any time derivative of the nondynamical modes). We solve them to express the nondynamical modes in terms of the dynamical ones. Inserting these expressions back into (\ref{genYNact}), we obtain the action for the dynamical modes. This action is formally of the type
\begin{equation}
S_{Y}^{(2)} = \frac{1}{2} \int d t \, d^3 k \left[ \dot{Y}_i^* \, K_{ij} \, \dot{Y}_j + \left( \dot{Y}_i^* \, X_{ij} \, Y_j +  {\rm h. \, c.} \right) - Y_i^* \, \Omega_{ij}^2 \, Y_j \right]
\label{genYact}
\end{equation}
where (up to boundary terms) the matrices $K$ and $\Omega^2$ are hermitian, while $X$ anti-hermitian. Extremizing this action we find
\begin{eqnarray}
\frac{\delta S_{Y}^{(2)}}{\delta Y_i^*} = 0 &\Rightarrow&
K_{ij} \, \ddot{Y}_j + \left( \dot{K}_{ij} + X_{ij} - X_{ji}^* \right) \dot{Y}_j + \left( \dot{X}_{ij} + \Omega_{ij}^2 \right) Y_j = 0 \nonumber\\
&\Rightarrow& \ddot{Y}_i = - \left[ K^{-1} \left( \dot{K} + X - X^\dagger \right) \right]_{ij} \dot{Y}_j - \left[ K^{-1} 
\left( \dot{X} + \Omega^2 \right) \right]_{ij} Y_j 
\label{gen-eq-form-2}
\end{eqnarray}
These two equations are precisely the linearized equations (\ref{Eq11}) and (\ref{Eq2}), after we have inserted in them the expressions for the nondynamical modes.

It is straightforwards to obtain the expressions of these matrices
from the terms in the action (\ref{action-2dS-acw}). However, they are
rather involved, and not needed for the present discussion.  We list
instead the expression for the determinant of the kinetic matrix
\begin{eqnarray}
&& {\rm det} \, K = \frac{p_T^6\, \left( 1 + \mu^2 \right)^2}{2\,
\Delta\, p_L^2\, p^2}\, \left[p_L^2 - H_a^2\,
\left(1+\mu^2\right)\, \left(2+\mu^2\right)\right] \nonumber\\
&& \Delta \equiv \left[2\, p^2 + \mu^2\, \left( p^2 + p_L^2
\right)\right]^2 -2 H_a^2\,
\left(1+\mu^2\right)\,\left(2+\mu^2\right)\, \left[ 2 p_L^2\,
(1+\mu^2)^2+p_T^2\,(2+\mu^2)\,(1+2\mu^2)\right] 
\label{detK-acw}
\end{eqnarray}
One can verify that the determinant starts positive at $p \gg H$, and becomes zero precisely when $p_L = \sqrt{2+\mu^2} \, \sqrt{H_a \, H_b} = \sqrt{2+\mu^2} \, \sqrt{1+\mu^2} \, H_a \,$. As it is clear from the general form (\ref{gen-eq-form-2}) of the equations, one expect that the solutions diverge at this moment. This is precisely what we have found by explicitly solving these equations in the previous Section !

We also see that the determinant of the kinetic matrix becomes negative after this moment. Therefore, one of the physical modes of the system becomes a ghost. This by itself signals an instability of the ACW vacuum.

\section{Simplified computation}
\label{sect-simplified}

In this Section we present a simplified stability analysis, which includes only the perturbations of the vector field. This analysis is not rigorous, since the metric perturbations are sourced by these perturbations already at the linearized level. However, as we now show, this simplified study clearly shows the origin of the instability without the involved algebra that is necessary to deal with the perturbations of the metric. We start from the ACW background, given in eqs.  (\ref{background}), 
(\ref{vector-back-acw}), and (\ref{Hab-acw}). The perturbations of the vector field are as in 
Section \ref{sect-linearized},
\begin{equation}
\delta A_\mu = \left( \alpha_0 ,\, \alpha_1 ,\, \partial_i \alpha \right)
\end{equation}
where we disregard the decoupled $2$d vector mode $\alpha_i \,$. 

The constraint equation enforced by the lagrange multiplier, once linearized in the perturbations, gives 
$\alpha_1 = 0 \,$ (we recall, that, when the metric perturbations were included, the same constraint enforced $\hat\alpha_1 = \mu\, \hat\Psi$).

The equations for the vector field (specifically, the second line of  (\ref{eqs-acw-nol}), in which the lagrange multiplier has been eliminated) then give
\begin{eqnarray}
&& p_T^2\, \dot\alpha - \left[ p_L^2 + p_T^2 - 2\, \left( 1+ \mu^2
\right)\, H_a^2 \right]\, \alpha_0 = 0 \nonumber\\
&& \ddot{\alpha} + H_a\, \dot\alpha - \dot{\alpha}_0 + \left[
p_L^2 - 2 \left( 1 + \mu^2 \right)\, H_a^2 \right]\, \alpha -
H_a\, \alpha_0 = 0 
\label{appacw:maxwell-2}
\end{eqnarray}

We proceed by differentiating the first equation with respect to time, and by
combining it with the second equation so to eliminate the $\ddot{\alpha}$ term. 
In this way we get two first order differential equations:
\begin{eqnarray}
\dot{\alpha} &=& \frac{p_L^2 + p_T^2 - 2 \left( 1 + \mu^2
\right)\, H_a^2}{p_T^2}\, \alpha_0 \nonumber\\
\dot{\alpha}_0 &=& - \frac{\left( 1 + 2\, \mu^2 \right)\, p_L^2 -
2\, \left( 1 + \mu^2 \right)\, \left( 3 + 2\, \mu^2 \right)\,
H_a^2}{p_L^2 - 2\, \left( 1 + \mu^2 \right)\, H_a^2}\, H_a\,
\alpha_0 - p_T^2\, \alpha \label{appacw:maxwell-2dS}
\end{eqnarray}
This set of equations has a singular point when $p_L^2 = 2\,
\left( 1 + \mu^2 \right)\, H_a^2 = 2 \, H_a \, H_b \,$. This is the 
divergence which corresponds to the one
occurring at $p_L^2 =  \left( 2 + \mu^2 \right)\,  H_a \, H_b \,$ 
in the full computation of the two previous Sections. 
We set the origin of time at the singularity, 
so that the physical momenta are
\begin{equation}
p_L = \sqrt{2 \left( 1 + \mu^2 \right)} \, H_a  \, {\rm e}^{-H_a \, t} \;\;\;,\;\;\;
p_T = p_{T0} \, {\rm e}^{-\left( 1 + \mu^2 \right) H_a \, t}
\end{equation}

Once expanded close to the singularity, the system (\ref{appacw:maxwell-2dS}) becomes
\begin{equation}
\dot{\alpha} \approx \alpha_0 \,\,\,\, , \,\,\,\,\, 
\dot{\alpha}_0 \approx -\frac{\alpha_0}{t} - p_{T0}^2\, \alpha
\end{equation}
which we can combine into (we obtain the same result if we first combine
the two equations (\ref{appacw:maxwell-2dS}) into a unique equations for $\alpha$, and we then expand that one)
\begin{equation}
\ddot{\alpha} + \frac{\dot\alpha}{t} + p_{T*}^2\, \alpha \approx 0
\label{appacw:maxwell2dS-apprx}
\end{equation}
which is in turns solved be
\begin{equation}
\alpha \approx C_1\, J_0 \left( p_{T*}\, t \right) + C_2\, Y_0
\left( p_{T*}\, t \right) \label{appacw:apprx-soln-2dS}
\end{equation}
where $C_{1,2}$ are constants to be determined from initial
conditions. While the $J_0$ solution is constant at $t=0$, the
$Y_0$ solution has a logarithmic divergence. In principle, 
one may imagine arranging the initial conditions, so that the solution
will be regular ($C_2 =0$) at $t=0$. One would need to do so for
both the real and imaginary parts of $\alpha$ and for all modes (namely, for
all values of the comoving momenta). There is however no physical reason
 for this tuning; the initial conditions are given when the mode is deeply inside the horizon,
well before the equations become singular (there is no reason why the mode should 
initially ``know'' about the singularity that will happen close to horizon crossing).
Moreover, as we will see, the solution is clearly divergent ($C_2 \neq 0$) if the initial 
conditions are chosen in the adiabatic vacuum. Therefore, $\alpha$ has a logarithmic divergence, and $\alpha_0 \approx \dot{\alpha}$ has a linear $1/t$ divergence
close to horizon crossing, which indicates that the
background solution is unstable. This result is confirmed by the
numerical solutions below. Moreover, this degree of divergence 
is the same as the one obtained in the
complete computation of Subsection~(\ref{subsect-linsol}).

To find the initial conditions, and to understand the reason for the instability, 
we compute the quadratic action for the perturbations. We find
(up to boundary terms)
\begin{eqnarray}
S_{s} &=& \frac{1}{2}\, \int d^3 k\, dt\, a\, b^2\, \Bigg\{
p_T^2\, \vert \dot\alpha \vert^2 - p_T^2\, \left( \alpha_0^*\,
\dot\alpha + {\rm h.c.} \right) - p_T^2\, \left[ p_L^2 - 2\,
\left( 1 + \mu^2 \right)\, H_a^2 \right]\, \vert \alpha \vert^2
\nonumber\\
&& \qquad\qquad\qquad\qquad + \left[ p_L^2 + p_T^2 - 2\, \left( 1
+ \mu^2 \right)\, H_a^2 \right]\, \vert \alpha_0 \vert^2 \Bigg\}
\label{appacw:action-2ds}
\end{eqnarray}
This actions leads to the equations (\ref{appacw:maxwell-2dS}),
when it is extremized. The mode $\alpha_0$ appears without any
time derivatives in the action, therefore it can be integrated out
by solving its equation of motion, which gives
\begin{equation}
\alpha_0 = \frac{p_T^2}{p_L^2 + p_T^2 - 2\, \left( 1 + \mu^2
\right)\, H_a^2}\, \dot\alpha
\label{appacw:a0sol}
\end{equation}
Inserting this solution back into the action we get
\begin{equation}
S_{s} = \frac{1}{2}\, \int d^3 k\, dt\, a\, b^2\, p_T^2\, \left(
p_L^2 - 2\, \left( 1 + \mu^2 \right)\, H_a^2 \right)\, \left[
\frac{\vert \dot\alpha \vert^2}{p_L^2 + p_T^2 - 2\, \left( 1 +
\mu^2 \right)\, H_a^2} - \vert \alpha \vert^2 \right]
\label{appacw:action-2ds-2}
\end{equation}
In the early-time/UV limit when $p_{L,T}^2 \gg H_a^2$, this action
is ghost free and $\alpha$ is stable. However, there is a moment
later in the evolution, close to horizon crossing, when the
longitudinal physical momentum becomes $p_L = \sqrt{2\, \left( 1 +
\mu^2 \right)}\, H_a$ and the action vanishes. This generates the singularity in 
the equations (\ref{appacw:maxwell-2dS}). 
In addition, $\alpha$ becomes a ghost after
the action vanishes, since the kinetic term becomes negative. 
\footnote{The kinetic term also diverges when $ p = \sqrt{2\, \left( 1 + \mu^2 \right)}\, H_a
$ (where $p = \sqrt{p_L^2 + p_T^2}$); this is due to the fact that $\alpha_0$
cannot be integrated at this point. However, this happens after the singularity we are interested in.}

The canonical variable of the system is
\begin{equation}
\alpha_c \equiv \sqrt{a}\, b\, p_T\, \sqrt{\frac{p_L^2 - 2\, \left( 1 +
\mu^2 \right)\, H_a^2}{p_L^2 + p_T^2 - 2\, \left( 1 + \mu^2
\right)\, H_a^2}}\, \alpha \label{appacw:alpha-can}
\end{equation}
Inserting this into the action (\ref{appacw:action-2ds-2}) we
get (up to a boundary term)
\begin{equation}
S_s = \frac{1}{2}\, \int d^3 k\, dt\, \left( \vert \dot{\alpha}_c
\vert^2 - \omega_c^2\, \vert \alpha_c \vert^2 \right) 
\label{appacw:action-2ds-can}
\end{equation}
The exact expression of $\omega_c$ is rather involved, and we do not report it here.
The first two terms in the early time expansion ($H \ll p$) are
\begin{equation}
\omega_c  = p \left[ 1 - \frac{\left( 9 + 8 \mu^2 \right) p_L^4 + \left(3 + 2 \mu^2 \right)^2 p_T^4 + 2 \left( 
9 + 10 \mu^2 - 4 \mu^4 \right) p_L^2 \, p_T^2}{8 p^4} \, \frac{H_a^2}{p^2} + {\rm O } \left( \frac{H_a^4}{p^4} \right) \right]
\end{equation}

The frequency $\omega_c$ is adiabatically evolving at early times, $\dot{\omega_c} \ll \omega_c = 
{\rm O } \left( H / p \right) \ll 1 \,$. Therefore, we can set the initial conditions for the canonical variable $\alpha_c$ in the adiabatic vacuum
\begin{equation}
\alpha_{\rm c,\, in} = \frac{1}{\sqrt{2 \, \omega_c}} \, {\rm e}^{- i\,\int^t \omega_c \, d t}
\end{equation}
This gives the initial values of $\alpha_c$ and $\dot{\alpha_c}$. From (\ref{appacw:action-2ds-can}) and its time derivative we then obtain the initial values of $\alpha$ and $\dot{\alpha} \,$. Finally, from 
(\ref{appacw:a0sol}) we obtain the initial value of $\alpha_0 \,$. Specifically, we find
\begin{equation}
{\rm Im } \left( \alpha_0 / \alpha \right)_{\rm in}  = - \frac{p_T^2}{p} \left[ 1 + {\rm O } \left( \frac{H^2}{p^2} \right) \right] \;\;\;,\;\;\;
{\rm Re } \left( \alpha_0 / \alpha \right)_{\rm in} = - \frac{\mu^2 \, p_T^4}{2 \, p^3} \, \frac{H_a}{p} \left[ 1 + {\rm O } \left( \frac{H^2}{p^2} \right) \right] 
\end{equation}

\begin{figure}[h]
\centerline{
\includegraphics[width=0.4\textwidth,angle=-90]{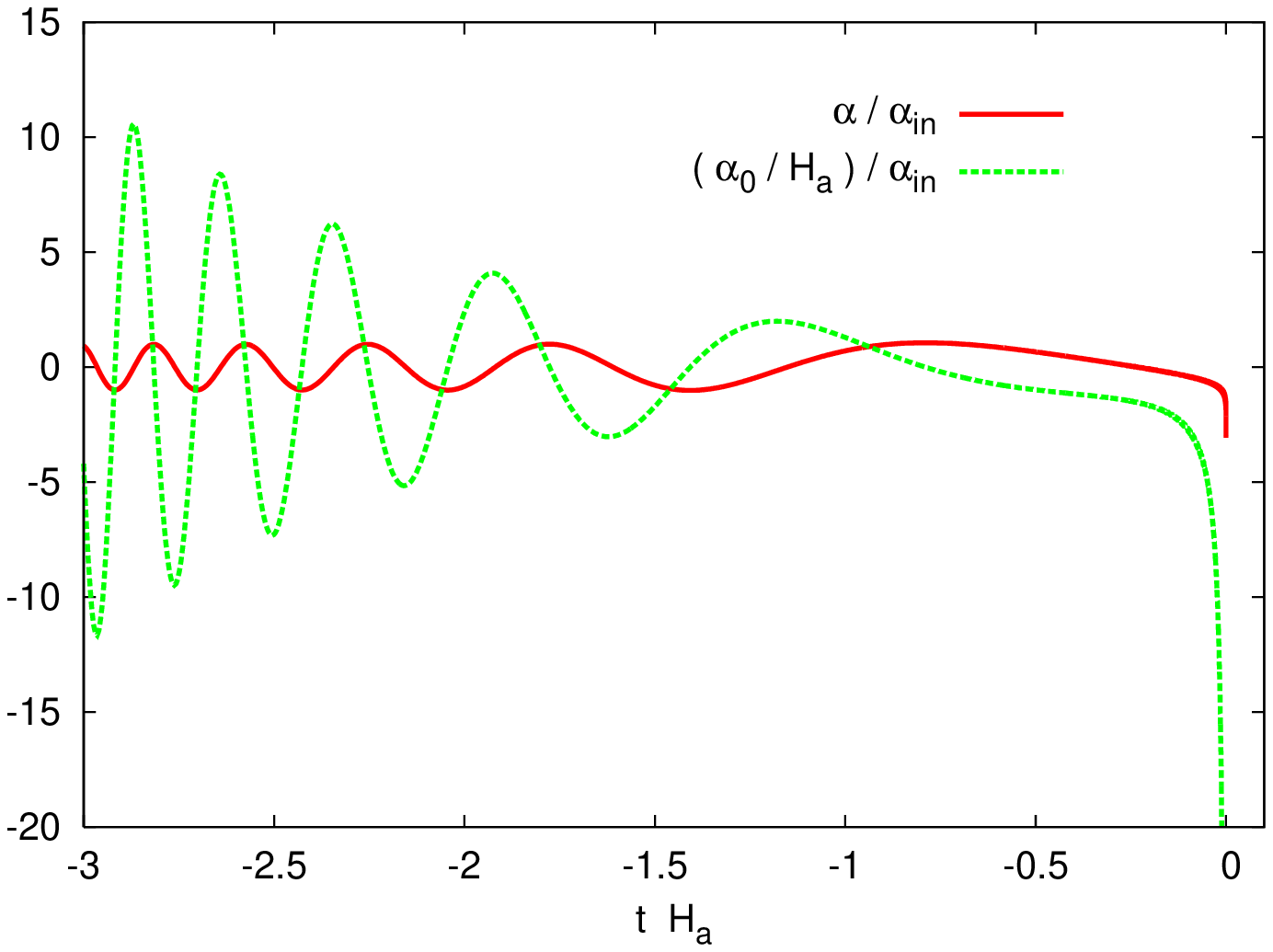}
\includegraphics[width=0.4\textwidth,angle=-90]{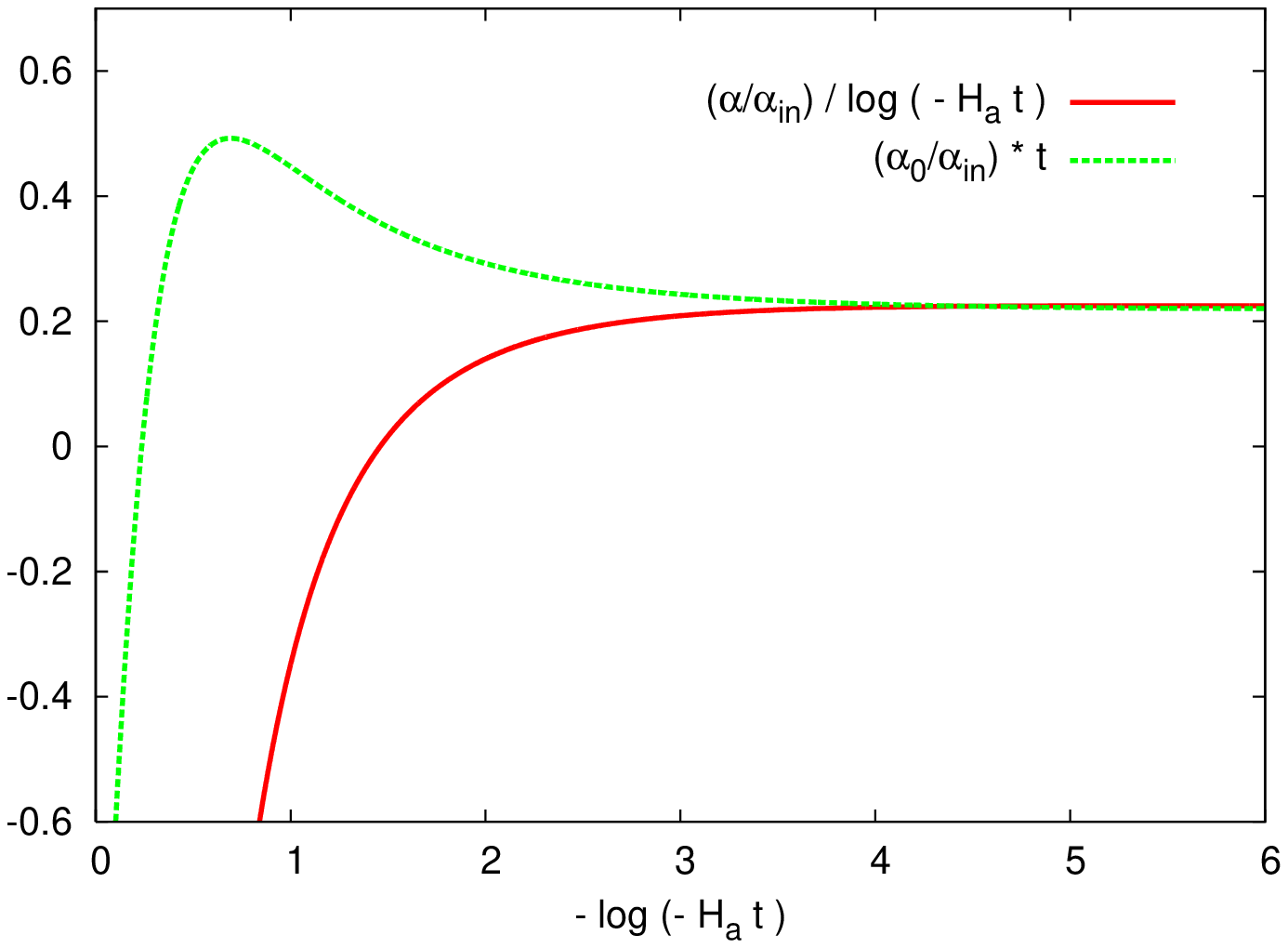}
} \caption{The right panel shows a zoom of the left panel in the
region $-1 \leq H_a \, t \leq - 2.5 \cdot10^{-3 } \,$ (the time is shown
in log units). Both modes blow up when the system
(\ref{appacw:maxwell-2dS}), becomes singular, $\alpha \propto {\rm
log } t$ and $\alpha_0 \propto 1/t \,$, in agreement with the
analytic approximate solution (\ref{appacw:apprx-soln-2dS}).
Notice also that $\alpha_0 \approx \dot{\alpha}$ close to the singularity,
in agreement with the analytic computation.}
\label{fig:fig2}
\end{figure}

The overall phase of the modes is unphysical, and we can always choose $\alpha_{\rm in}$ to be real.
Since the coefficients of (\ref{appacw:maxwell-2dS}) are real, we need to evolve this system twice, one for the real part and one for the imaginary parts of the modes. We did so starting from $t = - 7 / H_a \,$, 
so that $H/p = {\rm O}\left( 10^{-3} \right)$ initially. For definiteness, we take $\mu = 0.1 ,\, p_{T0} = H_a \,$ (as in the full computation leading to Figure \ref{fig:fig1}). We show the solutions for the real part in Figure \ref{fig:fig2}. The right panel is a close up near the singularity (which, in logarithmic units, occurs at $-\ln\left(-H_a\, t \right) = +\infty$). The two solutions have been re-scaled so to make the degree of divergence manifest. The imaginary part presents an analogous divergence.

\section{Appearance of  a ghost in other theories}

\label{sect-other}

We have performed a stability analysis also of the models
\cite{ford,mukhvect,soda1,Dimopoulos1}. The computations for these models are 
more involved than the one presented here, since they involve a larger
number of perturbations.~\footnote{A complete computation is
presumably even more complicated for models where the kinetic term
for the vector field is highly noncanonical, see for instance
Ref. \cite{Bamba:2008be}.} They will be reported elsewhere.
We anticipate that these models have a ghost close to horizon crossing. In addition,
for the two models \cite{mukhvect,soda1} the linearized equations blow up
close to horizon crossing, analogously to what happens in the ACW case.

The three models \cite{mukhvect,soda1,Dimopoulos1} are characterized by a nonminimal 
coupling of one or more vector fields to the curvature. The presence of the ghost is not due
to an anisotropic geometry (in fact, the background evolution can be taken isotropic in
\cite{mukhvect}, by having $3N$ mutually orthogonal vectors, and it is isotropic in 
\cite{Dimopoulos1}) but rather to the sign of the coupling to the curvature. The action for the vector field(s) is
\begin{equation}
S = \int d^4 x \sqrt{-g} \left[ - \frac{1}{4} F^2 + \frac{1}{12} \, R \, A^2 \right]
\label{act-vec-ghost1}
\end{equation}
where the $1/12$ coefficient allows for a slow roll of the vector in \cite{mukhvect,soda1} (see the discussion in \cite{hcp1}) and the production of a (nearly) scale invariant spectrum of perturbations in \cite{Dimopoulos1}.~\footnote{See \cite{soda2} for a similar study, based on a different model of vector fields.} For simplicity, we assume that the vector field has no vev, and the background is pure de Sitter. Using conformal time $\eta$,
\begin{equation}
S =  \int d \eta \, d^3 x \left[ - \frac{1}{4} F^2 - \frac{M^2}{2} \, A^2 \right]
\label{act-vec-ghost2}
\end{equation}
where $M^2 = - 2 \, a^2 \, H^2$, and where the indices are risen with the inverse of the Minkowski metric
$\eta_{\mu \nu} = {\rm diag } \left( -1 ,\, +1 ,\, +1 ,\, +1 \right)$.

The fact that the longitudinal vector (\ref{act-vec-ghost1}) is a ghost in this model depends on the sign of the mass term. At first, it is not immediate why a mass term, and not a kinetic term, controls the nature of a mode. The main reason is that the longitudinal degree of freedom is present only for $M^2 \neq 0 \,$, since, for vanishing mass, the action (\ref{act-vec-ghost1}) is U(1) invariant, and the vector has only the two transverse degrees of freedom. We present three arguments to show that $M^2 < 0$ leads to a ghost.~\footnote{We thank A. Vainshtein for suggesting us to use the first two arguments, and for discussions. We thank A. J. Tolley for suggesting us to use the second argument. The third argument was presented in our previous work \cite{hcp1}.} In all the discussion below, $M \equiv \vert M^2 \vert > 0$. Moreover, only for this Section, $k^2 \equiv k^\mu \, k_\mu \,,\,\, \vec{k}^2 \equiv k_i \, k_i$.

The most standard way to compute how many states there are in a theory, and what their nature is, is by computing the propagator. The quadratic lagrangian in (\ref{act-vec-ghost2}) can be cast in the form $(1/2) A^\mu \, P_{\mu \nu}^{-1}  \, A^\nu \,$, where
\begin{equation}
P_{\mu \nu} = - \frac{\eta_{\mu \nu} + \frac{k_\mu \, k_\nu}{M^2}}{k^2 + M^2}
\end{equation}
is the propagator. In general, the propagator needs to be diagonalized. However, we can choose a frame in which it is diagonal (clearly, the number and nature of physical modes does not depend on the frame). For a positive $M^2$, the pole is at $k^2 = - M^2 < 0 $, and we can go in the rest frame, where $k^\mu = - k_\mu = \left( M ,\, 0 ,\, 0 ,\, 0 \right) \,$. In this case, $- \left( \eta_{\mu \nu} + k_\mu k_\nu / M^2 \right) = {\rm diag } \left( 0 ,\, - 1 ,\, -1 ,\, -1 \right)$, indicating that the theory has three well behaved physical particles ($-1$ indicates a well behaved mode due to the signature; cf. the propagator for a scalar particle). For $M^2 < 0 ,\,$ we cannot go in the rest frame; however, we can choose a frame where the energy vanishes,
$k^\mu = k_\mu = \left( 0 ,\, 0 ,\, 0 ,\, M \right) \,$. In this case, $- \left( \eta_{\mu \nu} + k_\mu k_\nu / M^2 \right) = {\rm diag } \left( 1 ,\, -1 ,\, -1 ,\, 0 \right)$, indicating that one mode (the longitudinal vector) is a ghost.

The ghost also appears in the Stuckelberg formalism. For simplicity, here $M$ is treated has a constant (the time dependence of $M$ does not modify the quadratic kinetic term, but it complicates the diagonalization). If we redefine
\begin{equation}
A_\mu = B_\mu + \frac{1}{M} \partial_\mu \phi
\end{equation}
we promote the action (\ref{act-vec-ghost2}) to a gauge invariant action, with the symmetry
\begin{equation}
\phi \rightarrow \phi + \xi \;\;\;,\;\;\; B_\mu \rightarrow B_\mu - \frac{1}{M} \partial_\mu \phi
\end{equation}
The action (\ref{act-vec-ghost2}) is recovered in the unitary gauge, $\phi = 0$. But one can also choose a gauge in which $B_\mu$ is transverse, $\partial^\mu B_\mu = 0 \,$. In this gauge, the action of the system is
\begin{equation}
S =  \int d \eta \, d^3 x \left[ - \frac{1}{4} F \left( B \right)^2 - \frac{M^2}{2} \, B^2 \mp \frac{1}{2} \left( \partial \phi \right)^2 \right]
\label{act-vec-ghost3}
\end{equation}
where the field strength $F_{\mu \nu}$ does not contain $\phi$, and where the kinetic term of 
$\phi$ has opposite sign to $M^2 \,$. We stress that the two actions (\ref{act-vec-ghost2}) and 
(\ref{act-vec-ghost3}) describe the same theory in two different gauges. We again see that the longitudinal component $\phi$ is a normal field for $M^2 > 0$, and a ghost for $M^2<0\,$.

A third way to find the ghost is to decompose the vector as $A_\mu = \left( \alpha_0 ,\, \partial_i \alpha + \alpha_i \right) \,$, where $\partial_i \alpha_i = 0 \,$. The action (\ref{act-vec-ghost2}) then separates in two decoupled pieces,
\begin{eqnarray}
S &=& \int d \eta d^3 x \left\{ \frac{1}{2} \left[ \alpha_i'^2 - \left( \partial_i \alpha_j \right)^2 - M^2 \alpha_i^2 \right] + \frac{1}{2} \left[ \left( \partial_i \alpha' \right)^2 - 2 \partial_i \alpha' \partial_i \alpha_0 - M^2 \left( \partial_i \alpha \right)^2 + \left( \partial_i \alpha_0 \right)^2 + M^2 \, \alpha_0 \right] \right\} \nonumber\\
&=& \int d \eta d^3 k \left\{ \frac{1}{2} \left[ \vert \alpha_i' \vert^2 - \left( \vec{k}^2 + M^2 \right) \vert \alpha_i \vert^2 \right] + \frac{1}{2} \left[ \vec{k}^2 \vert \alpha' \vert^2 - \vec{k}^2 \left( \alpha'^* \, \alpha_0 + {\rm h. c.} \right) - M^2 \, \vec{k}^2 \vert \alpha \vert^2 + \left( \vec{k}^2 + M^2 \right) \vert \alpha_0 \vert^2 \right] \right\} \nonumber\\
\label{act-vec-ghost4}
\end{eqnarray}
where prime denotes derivative wrt conformal time, and in the second line we have Fourier transformed the modes as in~(\ref{Fourier}). The action splits in two decoupled parts. The first part governs the two transverse polarizations, which are well behaved. The second part contains only one physical mode, since $\alpha_0$ enters without time derivatives and must be integrated out. The equation of motion for this field obtained from (\ref{act-vec-ghost4}) is $\alpha_0 = \left[ \vec{k}^2 / \left( \vec{k}^2 + M^2 \right) \right] \alpha' \,$. Inserting this back into the second part of (\ref{act-vec-ghost4}), we obtain
\begin{equation}
S_{\rm longitudinal} = \int  d \eta  \, d^3 k \, \frac{\vec{k}^2 \, M^2}{2} \left[ \frac{\vert \alpha' \vert^2}{\vec{k}^2 + M^2} - \vert \alpha \vert^2 \right]\,.
\end{equation}
Again, we see that the longitudinal vector  is well behaved if $M^2>0$, and a ghost if $M^2 <0 \,$. As a check that this action is correct, we can compute the equations of motion enforced by it,
\begin{equation}
\alpha'' + \frac{2 k^2 a H}{k^2 + M^2} \alpha' + \left( k^2 + M^2 \right) \alpha = 0
\end{equation}
where we have used the fact that, for $M^2 = - 2 a^2 H^2$, one has $\left( M^2 \right)' = 2 a H M^2 \,$. This is precisely the equation of motion for the longitudinal vector already given in the literature, see for instance eq. (45) of \cite{Dimopoulos1} (notice that, in their convention, $R$ has the opposite sign wrt our convention).

\section{Discussion}
\label{sect-conclusion}

We have shown by explicit computations that the ACW background
solution is unstable. We did so in Section \ref{sect-linearized} by
writing down and solving the linearized equations for the most general
set of perturbations of the model. These equations, and their
solutions, diverge at some point $t_*$ close to horizon crossing (this
occurs for any mode in the Fourier decomposition; we stress that all
the modes are decoupled from each other at the linearized level). The
computations of Section \ref{sect-linearized} prove by themselves that
the background is unstable (it is possible that the divergency does
not occur at the full nonlinear level. However, when the nonlinear
interactions become important, the solution will be substantially
different from the ACW background solution). The computations the we have presented next
clarify the origin of the instability. Specifically, we
have shown in Section \ref{sect-ghost} that one dynamical variable of the
system becomes a ghost at $t_*$. The kinetic term for this mode
vanishes at $t_* \,$, which explains why the corresponding equation of
motion, and the solutions, diverge at this point. We stress that the
presence of a ghost indicates on its own an instability of the vacuum (see below). Our analysis is conceptually straightforward, but algebraically
very involved. For this reason, we have also shown that our expressions pass several
nontrivial checks. To gain a further understanding on the instability, we have also given, in
Section~\ref{sect-simplified}, a simplified stability study, including
only the perturbations of the vector field. The algebra involved is
now considerably simpler than in the previous case. We verified that this
system presents the same instability of the complete computation: the
solutions diverge when one mode becomes a ghost. As we mentioned in
the Introduction, the metric perturbations complicate the computation,
but do not remove the intrinsic instability of the model, which is
due to the vector field.

In a loose sense, the ACW model is ``doubly unstable'', since the linearized equations
for the perturbations diverge - we call this ``instability I'' - and since it has a ghost - ``instability II'' (in a generic model, either instability may appear without the other one). Motivated by this finding, we investigated whether other models with vector fields playing some role during inflation are also
unstable. To our knowledge, the fist model where anisotropic inflation is due to a vector is the one of Ford~\cite{ford}. The original proposal studied a subset of perturbations in the infinite wavelength limit.
A general computation, performed along the line of the one presented here, shows that this model contains a ghost in the short wavelength regime (it has the instability II mentioned above, but not the instability I). The models \cite{mukhvect,soda1,Dimopoulos1} are characterized by a nonminimal coupling of the vector to the curvature, with a precise factor. We have showed in Secton \ref{sect-other} why this leads to a ghost (the presence of a ghost is confirmed by the complete computation of the perturbations; we stress that the instability is due to the specific sign mass term, and not to the anisotropic expansion). The models \cite{mukhvect,soda1} present also the instability of the type I, as defined above, although this is not manifest. The detailed computations will be presented elsewhere.

An instability of the type I is a clear problem of the linearized theory. The implications of having a ghost instead require an understanding of the full theory. Such a study is beyond the goal of the present work. Nonetheless, we would like to provide some general comments. Although, contrary to what we have presented so far, these final considerations are not based on explicit computations, we believe that they are worth mention, and, possibly, more specific future investigations. 

The most obvious problem associated to a ghost is the instability of the vacuum. If a ghost is coupled to a normal field (and, in all the above theories, there are at least gravitational couplings), the vacuum will decay in ghost-nonghost excitations, with a rate which is UV divergent. This is simply due to phase space considerations. In normal situations, a source of any given energy can produce quanta up to some given momentum; this cuts-off the phase space available to the decay products. However, the quanta of a ghost field have negative energy: the higher their momentum, the more negative the energy is. Therefore, even a zero energy source decays into ghost-nonghost excitations conserving both energy and momentum. One may hope that the nonlinear interactions may somewhat arrange to cancel the 
total decay rate. However, this does not appear likely. The usual approach is to regard theories with ghosts as effective theories, valid only below some energy scale. Inflationary predictions heavily rely on the initial conditions; for instance, a slowly rolling inflaton field results in a nearly scale invariant spectrum of cosmological perturbations because the initial modes of the inflaton have amplitude $\vert \delta \phi \vert \propto k^{-1/2} \,$. This is due to the choice of an initial adiabatic vacuum in the quantized theory for the perturbations, which is performed in the deep UV regime (energy $\gg H^{-1}$) \cite{mfb}. In presence of a cut-off, the initial adiabatic vacuum cannot be imposed at arbitrarily early times, and, depending on the cut-off, it may not be possible to impose it at all. This casts doubts on the phenomenological predictions obtained for such models.

In fact, all these theories require a cut-off which makes them invalid
at high energies, irrespectively of the sign of the mass
term.~\footnote{Several of the considerations presented here emerged
  from discussions with N. Kaloper, who we thank for either pointing
  them out, or for stressing their relevance. We also acknowledge very
  useful discussions with D.H. Lyth on these issues.} We can see this
based on the behavior of massive vector fields at high energies. The
models studied here have a gauge invariance that is broken in a hard
way by the explicit mass term $M^2 \, A^2 \,$ for the vector. It is
well known that, in such cases, the interactions of the longitudinal
bosons violate unitarity at a scale which is parametrically set by
$M$, leading to a quantum theory out of control. For the present
models, $M$ is the Hubble rate or below, so that the entire
sub-horizon regime may be ill-defined. Although we are aware of
explicit computations of this problem only in Minkowski spacetime, we
believe that it applies also to the inflationary case, if one has
unbroken Lorentz invariance and transitions to a locally flat frame (moreover,
during inflation the mass term for the vector is nearly constant, and the 
momentum is adiabatically varying in the sub horizon regime); 
this issue is less transparent for the specific ACW model, in which
the constraint enforced by $\lambda$ will modify the interactions of the vector;
however, we believe that this concern is legitimate also for that
model. One may then adopt the approach of ignoring the ill behaved
longitudinal vector mode, and concentrate on the other degrees of
freedom in the model. We do not believe that this approach is
justified. At high energies, the longitudinal mode will also interact 
strongly with the other fields in the theory (this will renormalize the
coupling constants). Then, depending on exactly when this happens, the quantum
theory of the perturbations may be out of control throughout the
entire short wavelength regime. If this is the case, all initial conditions would become
unjustified, and the theory would lose its predictive power.

We are well aware that these problems, if indeed they are present, will
affect also the finding of the present work. However, in our opinion,
the computations presented above would not lose their importance. One
reason for this is that the instabilities that we have found have not
emerged so far in the literature. Different studies have adopted
different approximation schemes, which are very useful in the
case of scalar fields, as for instance studying separately the short
or long wavelength regimes \cite{dgw}, approximating the source with
fluids \cite{mota1} \footnote{Clearly, the microscopic description of
  the fluids considered in \cite{mota1} does not need to involve
  vector fields, although this has been considered as one of the
  possibilities in that work.}, or using the $\Delta N$ formalism
\cite{Dimopoulos1} (which allows to follow the evolution of the
perturbations only in the long wavelength regime). Although such
studies are perfectly legitimate in the regime of their validity, and
can also give very useful results in the present case, one should be
aware that these models require extra checks. We believe that our work
offers some nontrivial technical advance in this respect.

A more important reason is that ``curing'' a theory which has a hard
vector mass and a ghost is more problematic than curing a theory with
only a hard vector mass. The most immediate UV completion of a theory
with a hard mass is through a higgs mechanism. The mass would be then
due to the vev of a scalar field that becomes dynamical above the
scale $M$. In this way the theory remains under control also in the
short wavelength regime, and one can apply all the standard
computations valid for scalar fields during inflation. However, if
$M^2$ needs to have the wrong sign, the scalar field in this UV
completed theory needs to be a ghost. UV completing a theory with a
condensed ghosts may present analogous problems to the UV
completion~\cite{Cheng:2006us} of the ghost condensation model of
\cite{ArkaniHamed:2003uy}. We hope to return to this issue in some
future investigation.

\begin{acknowledgments}

We thank K. Dimopoulos, D.H. Lyth, A.J. Tolley, A. Vainshtein, 
M.B. Voloshin, and particularly N. Kaloper for very useful discussions. The work of
B.H. and M.P. was partially supported by the DOE grant
DE-FG02-94ER-40823.

\end{acknowledgments}

\bigskip

\appendix

\section{Explicit expressions for the computation of Subsection \ref{subsect-linsol}}

\label{appexplicit}

We write here some explicit steps in the linearized computation of the perturbations performed in
Subsection \ref{subsect-linsol}.

The explicit form of equations (\ref{pert-eqns-acw}) is 
\begin{eqnarray}
{\rm Eq}_{00}: && -\frac{1}{M_p}\, \Bigg\{ \left(2 + \mu^2
\right)\, H_a\, \dot{\hat\Psi} + p_T^2\, \hat\Psi + \left( 2 +
\mu^2\right)\, \left( 3 + 2\, \mu^2 \right)\, H_a^2\, \hat\Phi -
2\, \left(1+\mu^2\right)\, H_a\, \hat\chi \nonumber\\
&& \qquad\qquad - \left( 2 + \mu^2\right)\, H_a\, \hat{B} - \mu\,
H_a\, \hat\alpha_0 - \mu\, H_a\, \left( \dot{\hat\alpha}_1 - \mu\,
\dot{\hat\Psi} \right) - \mu\, H_a^2\, \left( \hat{\alpha}_1 -
\mu\, \hat\Psi \right) \Bigg\} = 0 \nonumber\\
{\rm Eq}_{01}: && -\frac{i\, p_L\, a}{M_p}\, \Bigg\{ 2\, \left( 1
+ \mu^2\right)\, H_a\, \hat\Phi + \frac{1}{2 p_L^2}\, \left[ p_T^2
- 4\, \mu^2\, \left(1+\mu^2\right)\, H_a^2\right]\, \hat\chi -
\frac{\hat{B}}{2} - 2\, \mu\, \left(1+\mu^2\right)\,
\frac{H_a^2}{p_L^2}\, \hat\alpha_0 \Bigg\}=0 \nonumber\\
{\rm Eq}_{0i} : && -\frac{i\, p_{Ti}\, b}{M_p}\, \left[
\dot{\hat\Psi} + \left( 2 + \mu^2 \right)\, H_a\, \hat\Phi -
\frac{\hat\chi}{2} + \frac{p_L^2}{2 p_T^2}\, \hat{B} - \mu\, H_a\,
\hat\alpha + \mu\, H_a\, \left( \hat\alpha_1 -\mu\, \hat\Psi
\right) \right] \nonumber\\
&& \qquad + \frac{b}{2}\, \left[ -p_L^2\, \dot{\hat{\tilde{B}}}_i
+ H_a\, p_L^2\, \left( 1+\mu^2\right)\, \hat{\tilde{B}}_i + p^2\,
\hat{B}_i - 2\, i \, \mu\, \frac{H_a}{b}\, p_L\,
\hat{\alpha}_i \right]=0 \nonumber\\
{\rm Eq}_{11}: && -\frac{\mu\, a^2}{M_p}\, \Bigg\{
\ddot{\hat\alpha}_1 + 2\,\left( 2 + \mu^2 \right)\, H_a\,
\dot{\hat\alpha}_1 + p_T^2\, \hat{\alpha}_1 + \dot{\hat\alpha}_0 +
\left( 3 + 2 \mu^2 \right)\, H_a\, \hat{\alpha}_0 - p_T^2\,
\hat\alpha - \mu\, H_a\, \dot{\hat\Psi} \nonumber\\
&& \qquad\qquad -\frac{2+\mu^2}{\mu}\, H_a\, \dot{\hat\Phi} +
\frac{1}{\mu}\, \left[ p_T^2 - \left( 2 + \mu^2\right)\, \left( 3
+ 2\mu^2\right)\, H_a^2\right]\, \hat\Phi + \frac{\dot{\hat
B}}{\mu}
+ \frac{3+2\mu^2}{\mu}\, H_a\, \hat{B} \nonumber\\
&& \qquad\qquad + \left( 9 + 8\, \mu^2\right)\, H_a^2\, \left(
\hat\alpha_1
- \mu\, \hat\Psi \right) \Bigg\}=0 \nonumber\\
{\rm Eq}_{1i} : && \frac{a\,b\, p_{Ti}}{2\, p_L\, M_p}\, \Bigg[
\dot{\hat\chi} + 3 H_a\, \hat\chi + \frac{p_L^2}{p_T^2}\, \left[
\dot{\hat{B}} + \left( 3 + 4\mu^2\right)\, H_a\, \hat{B} \right] -
2\,\mu \, H_a\, \left( \dot{\hat\alpha} + 3 H_a\, \hat\alpha
\right) + 2 p_L^2\, \hat\Phi \nonumber\\
&& \qquad\qquad -4\, \mu^3\, H_a^2\, \hat\alpha - 2\, \mu\, H_a\,
\hat\alpha_0 \Bigg] \nonumber\\
&& \qquad -\frac{i\, a\, b\, p_L}{2}\, \Bigg[
\ddot{\hat{\tilde{B}}}_i + \left( 1 + 2\, \mu^2\right)\, H_a\,
\dot{\hat{\tilde{B}}}_i - \dot{\hat{B}}_i - \left( 2 +
3\mu^2\right)\, H_a\, \hat{B}_i \nonumber\\
&& \qquad\qquad\qquad\quad + \left[ p_T^2 - H_a^2\,
\left(1+\mu^2\right)\, \left(2+3\mu^2\right) \right]\,
\hat{\tilde{B}}_i + 2\,i \, \frac{H_a\, \mu}{b\, p_L}\left(
\dot{\hat\alpha}_i + 2\, H_a\, \left(1+\mu^2\right)\, \hat\alpha_i
\right)
\Bigg]=0 \nonumber\\
{\rm Eq}_{ij}: && \frac{b^2}{M_p}\, \Bigg\{ \Bigg[ \ddot{\hat\Psi}
+ \left( 3 + 2\mu^2\right)\, H_a\, \dot{\hat\Psi} + p_T^2\,
\hat\Psi + \left( 2 + \mu^2\right)\, H_a\, \dot{\hat\Phi} - \left[
p^2 - H_a^2\, \left(2+\mu^2\right)\,
\left(3+2\mu^2\right)\right]\, \hat\Phi
-\dot{\hat\chi} \nonumber\\
&& \qquad\quad -\dot{\hat{B}} - \left( 3 + 2\, \mu^2\right)\,
H_a\, \hat{B} -\left( 3+\mu^2\right)\, H_a\, \hat\chi + \mu\,
H_a\, \dot{\hat\alpha}_1 + \mu\, H_a\, \hat{\alpha}_0 + \mu\,
H_a^2\, \left( \hat{\alpha}_1 - \mu\, \hat\Psi \right)  \Bigg]\, \delta_{ij}
\nonumber\\
&& \qquad\quad 
-p_{Ti}\, p_{Tj}\, \left[ \hat\Psi - \hat\Phi - \frac{1}{p_T^2}\,
\left( \dot{\hat{B}} + \left(3 + 2\mu^2\right)\,
H_a\, \hat{B} \right) \right] \Bigg\} \nonumber\\
&& \qquad\quad + i\, b^2\, \left[ p_{(Ti}\, \dot{\hat{B}}_{j)} +
\left( 2 + \mu^2 \right)\, H_a\, p_{(Ti}\, \hat{B}_{j)} + p_L^2\,
p_{(Ti}\, \hat{\tilde{B}}_{j)} \right] = 0 \nonumber\\
%
%\end{eqnarray}
%\begin{eqnarray}
{\rm Eq}_0: && -\frac{i\, p_T^2}{p_L}\, \Bigg\{ \dot{\hat\alpha} +
\frac{p_L^2}{p_T^2}\, \hat{\alpha}_1 + H_a\, \hat\alpha +
\frac{1}{p_T^2}\, \left[ p^2 - 2\, \left(1+\mu^2\right)\,
H_a^2\right]\, \hat{\alpha}_0 + \mu\, \frac{p_L^2}{p_T^2}\, H_a\,
\hat\Phi \nonumber\\
&& \qquad\qquad -2\, \mu\, \left( 1 + \mu^2\right)\,
\frac{H_a^2}{p_T^2}\, \hat\chi + \frac{p_L^2}{p_T^2}\, H_a\,
\left( \hat{\alpha}_1 - \mu\, \hat\Psi \right) \Bigg\} = 0
\nonumber\\
{\rm Eq}_1: && \frac{4\left(1+\mu^2\right)\, H_a^2}{a}\, \left(
\hat{\alpha}_1 - \mu\, \hat\Psi \right) = 0 \nonumber\\
{\rm Eq}_i: && = -\frac{p_{Ti}}{b\, p_L}\, \Bigg[
\ddot{\hat\alpha} + 3\, H_a\, \dot{\hat\alpha} +
\dot{\hat\alpha}_0 + \left( p_L^2 - 2\, \mu^2\, H_a^2 \right)\,
\hat\alpha - p_L^2\, \hat\alpha_1 + \mu\, \frac{p_L^2}{p_T^2}\,
H_a\, \hat{B} + 2\, H_a\, \hat\alpha_0 \Bigg] \nonumber\\
&& \qquad\qquad +\frac{i\, M_p}{b}\, \Bigg[ p_L\, \mu\, H_a\,
\left( \dot{\hat{\tilde{B}}}_i - \left( 1+\mu^2\right)\, H_a\,
\hat{\tilde{B}}_i - \hat{B}_i \right) - \frac{i}{b}\, \left[
\ddot{\hat\alpha}_i + H_a\, \dot{\hat\alpha}_i + \left( p^2 - 2\,
H_a^2\, \left(1+\mu^2\right)\right)\, \hat\alpha_i \right]
\Bigg]=0 \nonumber\\ \label{pert-eqns-acw-exp}
\end{eqnarray}
where the notation on the momenta is explained after eq.(\ref{Fourier}), and where
$x_{(i}\, y_{j)}\equiv \frac{1}{2} \left( x_i\, y_j + x_j\, y_i \right)$ denotes symmetrization. Notice that it has been possible to write all the above equations in terms of gauge invariant combinations. This is a nontrivial check on our algebra.

We recall that the indices $i,j = 1,2$ span the symmetric $y-z$ space. Equations carrying these indices contain both $2$d scalar and $2$d vector modes. However, it is easy to see that each of these equations separates in two independent equations, one for the $2$d scalars and one for the $2$d vectors: the equations in (\ref{pert-eqns-acw-exp}) carrying an $i$ index have the following structure
\begin{eqnarray}
{\rm Eq}_{0i} = k_{Ti} \, {\cal S}_1 + {\cal V}_{1i} \;,\;\;\;
{\rm Eq}_{1i} = k_{Ti} \, {\cal S}_2 + {\cal V}_{2i} \;,\;\;\;
{\rm Eq}_{ij} = \delta_{ij} \, {\cal S}_3 + k_{Ti} \, k_{Tj} \, {\cal S}_4 + k_{(Ti}\, {\cal V}_{3j)} \;,\;\;\;
{\rm Eq}_i = k_{Ti} \, {\cal S}_5 + {\cal V}_{4i}
\label{scaveeq}
\end{eqnarray}
where the expressions ${\cal S}_{1,\dots,5}$ contain the $2$d scalar, and the expressions 
${\cal V}_{1,\dots,4}$ contain the $2$d vector modes. Due to the transversality conditions of the $2$d vectors, 
\begin{equation}
k_{Ti} \, {\cal V}_{1i} = \dots = k_{Ti} \, {\cal V}_{4i} = 0 \;\;\;\Rightarrow\;\;\; {\cal V}_{1i} = \dots= {\cal V}_{4i} = 0
\end{equation}
so that we see explicitly that the equations for the $2$d scalars and the $2$d vectors are indeed decoupled from each other. We have verified that the system of $2$d vectors does not contain instabilities. For brevity, we disregard these modes in the following, and we concentrate only on the $2$d scalars.

We note that not all the equations in the system (\ref{pert-eqns-acw-exp}) are independent. Indeed, the Bianchi identities can be linearized to give
\begin{equation}
\nabla_{\mu}\left[G_{\nu}^{\mu} - \frac{1}{M_p^2}\, T_{\nu}^{\mu}
\right] = 0 \,\,\, \rightarrow \,\,\, g^{(0)\, \mu\alpha}\,
\nabla_{\mu}^{(0)}\, {\rm Eq}_{\alpha\nu} = 0
\end{equation}
where $\nabla^{(0)}$ denotes the covariant derivative constructed
from the background metric $g_{\mu\nu}$. These can be written as explicit
relations between the linearized equations (\ref{pert-eqns-acw})
\begin{eqnarray}
\nu=0 : && \frac{d}{dt}\, {\rm Eq}_{00} + \left( 3 + 2
\mu^2\right)\, H_a\, {\rm Eq}_{00} + i\, \left( \frac{p_L}{a}\,
{\rm Eq}_{01} + \frac{p_{Ti}}{b}\, {\rm Eq}_{0i}\right) +
\frac{H_a}{a^2}\, {\rm Eq}_{11} + \frac{\left(1+\mu^2\right)\,
H_a}{b^2}\, \delta^{ij}\, {\rm Eq}_{ij} = 0 \nonumber\\
\nu=1: && \frac{d}{dt}\, {\rm Eq}_{01} + \left( 3 + 2\mu^2
\right)\, H_a\, {\rm Eq}_{01} + i\, \left( \frac{p_L}{a}\, {\rm
Eq}_{11} + \frac{p_{Ti}}{b}\, {\rm Eq}_{1i} \right) = 0 \nonumber\\
\nu=i: && \frac{d}{dt}\, {\rm Eq}_{0i} + \left( 3 + 2\mu^2
\right)\, H_a\, {\rm Eq}_{0i} + i\, \left( \frac{p_L}{a}\, {\rm
Eq}_{1i} + \frac{p_{Tk}}{b}\, {\rm Eq}_{ki} \right) = 0
\label{bianchi-acw}
\end{eqnarray}
We explicitly checked that the system of equations
(\ref{pert-eqns-acw}) satisfies the above identities (which is a further nontrivial
check on our algebra). Therefore,
(\ref{bianchi-acw}) verifies that some of the linearized 
equations are redundant. By inserting the decomposition
(\ref{scaveeq}) into these identities (disregarding the $2$d vector parts
of the equations), we see that the $i=2,3$ identities become equal to each other.
We also see that we can use the three nontrivial identities 
($\nu=0,\,1,\,2$) to express ${\cal S}_2 ,\,$, 
${\cal S}_3$ and ${\cal S}_4$ in terms of the other equations.
Therefore, the ${\rm Eq}_{1i}$ and ${\rm Eq}_{ij}$ equations
in (\ref{pert-eqns-acw-exp}) can be obtained from the other ones, and
can be disregarded (obviously, one could equivalently choose to disregard
some other equations, as long as it is possible to express them in terms
of the remaining ones in (\ref{bianchi-acw})).

This leads us to eqs. (\ref{Eq1})-(\ref{Eq2}) of the main text. 
We solve equations (\ref{Eq01}) and (\ref{Eq02}) for $\hat\chi$ and $\hat{B}$.
The other equations become
\begin{eqnarray}
&&\frac{1}{H_a} \left[ - \frac{1}{\mu^2} \, \frac{p_T^2 \,
p^2}{p_L^2} + \frac{p_T^4}{2 \left( 1 + \mu^2 \right) p_L^2} +
\left(2 + \mu^2\right) \frac{p_L^2+2 \, p_T^2}{p_L^2} H_a^2
\right] \dot{\hat \Psi}
+ p_T^2 \, {\hat \Psi} \nonumber\\
&&\quad\quad+ \left[ - \frac{2 \left( 1 + \mu^2 \right)}{\mu^2} \,
p^2 - \frac{2 \, p_T^2}{\mu^2} + \left( 2 + \mu^2 \right)^2
\frac{p_T^2}{p_L^2} \left( - \frac{p_T^2}{2 \, \mu^2 \left( 1 +
\mu^2 \right)} + 2 \, H_a^2 \, \frac{p^2}{p_T^2} \right) -
\left( 2 + \mu^2 \right) H_a^2 \right] {\hat \Phi}  \nonumber\\
&&\quad\quad+ \frac{1}{\mu} \left[ \frac{p_T^2 \, p^2}{p_L^2} -
\mu^2 \, \frac{p_T^2}{p_L^2} \left( \frac{p_T^2}{2 \left( 1 +
\mu^2 \right)} + 2 \left( 2 + \mu^2 \right) H_a^2 \right) \right]
{\hat \alpha} + \left( 2 + \mu^2 \right) H_a \, \frac{p^2}{p_L^2}
\, \frac{{\hat \alpha}_0}{\mu} = 0 \nonumber\\ \label{Eq00-2} 
\end{eqnarray}

\begin{eqnarray}
&&\left[ \mu^2 + \frac{p_T^2}{p_L^2} \, \frac{p_T^2-4 \, \mu^2
\left( 1 + \mu^2 \right) H_a^2}{2 \, \mu^2 \left( 1 + \mu^2
\right) H_a^2} \right] \ddot{\hat \Psi} +\left[ \mu^2 \left( 3 + 2
\, \mu^2 \right) + \frac{p_T^2}{p_L^2} \left( \frac{1-2 \,
\mu^2}{\mu^2 \left( 1 + \mu^2 \right)} \,
\frac{p_T^2}{2 \, H_a^2} - 6 \right) \right] H_a \dot{\hat \Psi} + \mu^2 \, p_T^2 \, {\hat \Psi} \nonumber\\
&&\quad\quad+\left[  2 + \mu^2 + \frac{p^2}{p_L^2} \left[ - 2
\left( 2 + \mu^2 \right) + \frac{p_T^2}{\mu^2 \, H_a^2} \right] -
\frac{p_T^2}{p_L^2} \, \frac{p_T^2}{2 \left( 1 + \mu^2 \right)
H_a^2} \right]
H_a \dot{\hat \Phi} \nonumber\\
&&\quad\quad+\left[ \frac{1 + \mu^2}{\mu^2} + \frac{\left( 2 +
\mu^2 \right) \left( 1 - 2 \, \mu^2 \right)}{2 \, \mu^2 \left( 1 +
\mu^2 \right)} \, \frac{p_T^2}{p_L^2} -\left(2+\mu^2\right)\,
\left( \frac{3+2\mu^2}{p_T^2}
+ \frac{6}{p_L^2} \right)\, H_a^2 \right] p_T^2 \, {\hat \Phi} \nonumber\\
&&\quad\quad- \frac{1}{2 \, \mu \, H_a} \, \frac{p_T^2}{p_L^2}
\left( \frac{p_T^2}{1+\mu^2} - 4 \, \mu^2 \, H_a^2 \right)
\dot{\hat \alpha} + \mu \, \frac{p_T^2}{p_L^2} \left( 6 \, H_a^2 -
p_L^2 - \frac{1 - 2 \, \mu^2}{2 \, \mu^2 \left( 1 + \mu^2 \right)} \, p_T^2 \right) {\hat \alpha} \nonumber\\
&&\quad\quad- \left( \frac{p_T^2}{p_L^2} - \mu^2 \right)
\frac{\dot{\hat \alpha}_0}{\mu} - \left( 3 \, \frac{p_T^2}{p_L^2}
- 3 \, \mu^2 - 2 \, \mu^4 \right) H_a \, \frac{{\hat
\alpha}_0}{\mu} = 0 \nonumber\\ \label{Eq11-2} \\
&&-\frac{1}{\mu^2} \left( 1 - \mu^2 \, \frac{p_L^2}{p_T^2} \right)
\dot{\hat \Psi} - \frac{2 + \mu^2}{\mu} \, \frac{p^2}{p_T^2} \,
H_a \, {\hat \Phi} + \dot{\hat \alpha} + 2 \, H_a \, {\hat \alpha}
+ \frac{p^2}{p_T^2}
\, {\hat \alpha}_0 = 0 \nonumber\\ \label{Eq0-2} \\
&&\ddot{\hat \alpha} + 3 \, H_a \, \dot{\hat \alpha} + \left[
p_L^2 - \frac{p_T^2}{2 \left( 1 + \mu^2 \right)} \right] {\hat
\alpha} + \dot{\hat \alpha}_0 + H_a \, {\hat \alpha}_0 \nonumber\\
&&\quad\quad+\frac{1}{2 \, \mu \, H_a} \left(
\frac{p_T^2}{1+\mu^2} - 4 \, \mu^2 \, H_a^2 \right) \dot{\hat
\Psi} - \mu \, p_L^2 \, {\hat \Psi} + \frac{1}{\mu} \left[ p^2 -
\frac{\mu^2}{2 \left( 1 + \mu^2 \right)} \, p_T^2
- 2  \left( 2 + \mu^2 \right) \mu^2 \,  H_a^2 \right] {\hat \Phi} = 0 \nonumber\\
\label{Eq2-2}
\end{eqnarray}

We choose not to eliminate any more modes, but to rewrite these
equations as a system of differential equations for $\ddot{\hat \Psi}
,\, \ddot{\hat \alpha} ,\, \dot{\hat \Phi}$ and ${\hat \alpha}_0
\,$. The differential equations for $\hat{\alpha}_0$ and $\hat{\Phi}$
are obtained by differentiating the two equations (\ref{Eq00-2}) and
(\ref{Eq0-2}). In solving the numerical system we can replace
(\ref{Eq00-2}) and (\ref{Eq0-2}) with their time derivatives, provided
that these two equations are imposed as initial conditions (see
below).  We are thus left with the system of eqs. $\{
(\ref{Eq00-2})^{\bullet}, \, (\ref{Eq11-2}), \,
(\ref{Eq0-2})^{\bullet}, \, (\ref{Eq2-2}) \}$, where $({}^\bullet)$
indicates that we take the time derivative of the equation. This
system can be recast in the form (\ref{matrix}), in terms of the
coefficients
\begin{eqnarray}
\kappa_{22} &=& - \frac{1}{\mu} \left( \frac{p_T^2}{p_L^2} - \mu^2 \right) \nonumber\\
\kappa_{23} &=& \mu^2 + \frac{p_T^2}{p_L^2} \, \frac{p_T^2 - 4 \,
\mu^2 \left( 1 + \mu^2 \right) H_a^2}{2 \, \mu^2 \left( 1 + \mu^2 \right) H_a^2} \nonumber\\
\kappa_{24} &=& \left[  2 + \mu^2 + \frac{p^2}{p_L^2} \left[ - 2
\left( 2 + \mu^2 \right) + \frac{p_T^2}{\mu^2 \, H_a^2} \right] -
 \frac{p_T^2}{p_L^2} \, \frac{p_T^2}{2 \left( 1 + \mu^2 \right) H_a^2} \right] H_a \nonumber\\
\kappa_{32} &=& \frac{p^2}{p_T^2} \nonumber\\
\kappa_{33} &=& - \frac{1}{\mu} \left( 1 - \mu^2 \, \frac{p_L^2}{p_T^2} \right) \nonumber\\
\kappa_{34} &=& - \frac{ 2 + \mu^2 }{ \mu^2 } \, \frac{ p^2 }{ p_T^2 } \, H_a \nonumber\\
\kappa_{42} &=& \frac{ 2 + \mu^2 }{ \mu^2 } \, \frac{ p^2 }{ p_L^2 } \, H_a \nonumber\\
\kappa_{43} &=& \frac{p_T^2}{p_L^2} \, \frac{1}{H_a} \left[ -
\frac{p^2}{\mu^2} + \frac{p_T^2}{2 \left( 1 + \mu^2 \right)} +
\left( 2 + \mu^2 \right) \,
\frac{p_L^2 + 2 \, p_T^2}{p_T^2} \, H_a^2 \right] \nonumber\\
\kappa_{44} &=& - \frac{2 \left( 1 + \mu^2 \right)}{\mu^2} \, p^2
- \frac{2 \, p_T^2}{\mu^2} + \left( 2 + \mu^2 \right)^2 \,
\frac{p_T^2}{p_L^2} \left( - \frac{p_T^2}{2 \, \mu^2 \left( 1 +
\mu^2 \right)} + 2 \, H_a^2 \, \frac{p^2}{p_T^2} \right) - \left(
2 + \mu^2 \right) H_a^2 \nonumber\\
\label{coeffk}
\end{eqnarray}
and
\begin{eqnarray}
f_1 &=& - 3 \, H_a \, \dot{\hat \alpha} - \left[ p_L^2 -
\frac{p_T^2}{2 \left( 1 + \mu^2 \right)} \right] {\hat \alpha} - H_a \, {\hat \alpha}_0 \nonumber\\
&&\quad\quad-\frac{1}{2 \, \mu \, H_a} \left(
\frac{p_T^2}{1+\mu^2} - 4 \, \mu^2 \, H_a^2 \right) \dot{\hat
\Psi} + \mu \, p_L^2 \, {\hat \Psi} - \frac{1}{\mu} \left[ p^2 -
\frac{\mu^2}{2 \left( 1 + \mu^2 \right)} \, p_T^2
- 2  \left( 2 + \mu^2 \right) \mu^2 \,  H_a^2 \right] {\hat \Phi} \nonumber\\
f_2 &=&  -\left[ \mu^2 \left( 3 + 2 \, \mu^2 \right) +
\frac{p_T^2}{p_L^2} \left( \frac{1-2 \, \mu^2}{\mu^2 \left( 1 +
\mu^2 \right)} \, \frac{p_T^2}{2 \, H_a^2} - 6 \right) \right]
H_a \dot{\hat \Psi} - \mu^2 \, p_T^2 \, {\hat \Psi} \nonumber\\
&&\quad\quad-\left[ \frac{1 + \mu^2}{\mu^2} + \frac{\left( 2 +
\mu^2 \right) \left( 1 - 2 \, \mu^2 \right)}{2 \, \mu^2 \left( 1 +
\mu^2 \right)} \, \frac{p_T^2}{p_L^2} - \left(2+\mu^2\right)\,
\left( \frac{3+2\mu^2}{p_T^2} + \frac{6}{p_L^2} \right)\,
H_a^2 \right] p_T^2 \, {\hat \Phi} \nonumber\\
&&\quad\quad+ \frac{1}{2 \, \mu \, H_a} \, \frac{p_T^2}{p_L^2}
\left( \frac{p_T^2}{1+\mu^2} - 4 \, \mu^2 \, H_a^2 \right)
\dot{\hat \alpha} - \mu \, \frac{p_T^2}{p_L^2} \left( 6 \, H_a^2 -
p_L^2 - \frac{1 - 2 \, \mu^2}{2 \, \mu^2 \left( 1 + \mu^2 \right)}
\, p_T^2 \right) {\hat \alpha} \nonumber\\
&&\quad\quad + \left( 3 \, \frac{p_T^2}{p_L^2} - 3 \, \mu^2 - 2 \,
\mu^4 \right)
H_a \, \frac{{\hat \alpha}_0}{\mu} \nonumber\\
f_3 &=& - 2 \, H_a \, \dot{\hat \alpha} - 2 \, \mu^2 \, H_a \,
\frac{p_L^2}{p_T^2} \, {\hat \alpha}_0 - 2 \, \mu^3 \, H_a \,
\frac{p_L^2}{p_T^2} \, \dot{\hat \Psi} + 2 \left( 2 + \mu^2
\right) \mu \, H_a^2 \,
\frac{p_L^2}{p_T^2} \, {\hat \Phi} \nonumber\\
%
%\end{eqnarray}
%\begin{eqnarray}
f_4 &=& - \frac{1}{\mu} \, \frac{p_T^2}{p_L^2} \left[ p^2 - \mu^2
\left( \frac{p_T^2}{2 \left( 1 + \mu^2 \right)} +
2 \left( 2 + \mu^2 \right) H_a^2 \right) \right] \dot{\hat \alpha}  \nonumber\\
&&- \frac{p_T^2}{p_L^2} \left[ 3 \, p^2 - \frac{\mu^2 \, p_T^2}{1
+ \mu^2} + \frac{2 \, p^2}{\mu^2} - 4 \left( 2 + \mu^2 \right)
\mu^2 \, H_a^2 \right] \dot{\hat \Psi} +
2 \left( 2 + \mu^2 \right) \mu \, H_a^2 \, \frac{p_T^2}{p_L^2} \, {\hat \alpha}_0 \nonumber\\
&&+ \frac{H_a \, p_T^2}{p_L^2} \left[ \frac{2}{\mu} \, \left( 1 +
\mu^2 \right) p^2 + \frac{\mu}{1 + \mu^2} \, p_T^2 - 4 \left( 2 +
\mu^2 \right) \mu^3 \, H_a^2 \right] {\hat \alpha} +
2 \left( 1 + \mu^2 \right) H_a \, p_T^2 \, {\hat \Psi} \nonumber\\
&&- \frac{p_T^2}{p_L^2} \, \frac{H_a \left( 1 + \mu^2
\right)}{\mu^2} \left\{ \frac{1}{p_T^2} \left[ 2 \, p_L^2 + \left(
2 + \mu^2 \right) p_T^2 \right]^2 - \frac{\mu^4 \left( 2 + \mu^2
\right)^2}{\left( 1 + \mu^2 \right)^2} \left[ p_T^2 + 4 \left( 1 +
\mu^2 \right) H_a^2 \right] \right\} {\hat \Phi} \nonumber\\
\label{coefff}
\end{eqnarray}

As we already mentioned, we need to impose Eqs. (\ref{Eq00-2}) and (\ref{Eq0-2}) as
initial condition (this allowed us to replace these two equations with their time derivatives). 
By imposing them, we obtain the initial conditions for the nondynamical modes in terms of the dynamical ones:
\begin{eqnarray}
{\hat \alpha}_{0,in} &=& \frac{1}{p^2 \, {\cal D}_{in}} \Bigg\{ -
\Bigg[ 4 \left( 1 + \mu^2 \right) p_L^2 +
\left( 2 + \mu^2 \right)^2 p_T^2 \left( \frac{p_T^2}{1 + \mu^2 } - 4 \, \mu^2 \, H_a^2 \right) \nonumber\\
&& \quad\quad\quad\quad\quad\quad\quad\quad+ 2 \left( 2 + \mu^2
\right) p_L^2 \left( 2 \, p_T^2 - \left( 3 + 2 \, \mu^2 \right)
\mu^2 \, H_a^2 \right) \Bigg] \left( \dot{\hat \alpha} + H_a \,
{\hat \alpha} \right)
\nonumber\\
&&\quad\quad\quad- \,  \mu \, p_L^2 \left[ 2 \, p^2 -
\frac{\mu^2}{ 1 + \mu^2 } \, p_T^2 + 2 \left( 2 + \mu^2 \right)
H_a^2 \right] \left[ \mu \, H_a \, {\hat \alpha}
+ \left( 1 + \mu^2 \right) \left( 2 \, \frac{p^2}{p_T^2} - 1 \right) \dot{\hat \Psi} \right] \nonumber\\
&&\quad\quad\quad+ 2 \, \mu \left( 2 + \mu^2 \right) H_a \, p_L^2
\, p^2 \left[ \frac{2 \, H_a \left( 2 + \mu^2 \right) \left( 1 +
\mu^2 \right)}{p_T^2} \,
\dot{\hat \Psi} + {\hat \Psi} \right] \Bigg\}_{in} \nonumber\\
{\hat \Phi}_{in} &=& \frac{1}{{\cal D}_{in}} \Bigg\{ \left[ 2 \,
p^2 - \frac{\mu^2}{ 1 + \mu^2 } \, p_T^2 - 2 \, H_a^2 \left( 2 +
\mu^2 \right) \left( 1 + 2 \, \mu^2 \right) \right]
\left( - \frac{\dot{\hat \Psi}}{H_a} + \mu \, {\hat \alpha} \right) \nonumber\\
&&\quad\quad\quad+ 2 \, \mu^2 \, p_L^2 \, {\hat \Psi} - 2 \,
\mu \left( 2 + \mu^2 \right) H_a \left( \dot{\hat \alpha} + H_a \, {\hat \alpha} \right) \Bigg\}_{in} \nonumber\\
{\cal D} &\equiv& \frac{1+\mu^2}{p_T^2} \left\{ \left( 2 \, p^2 -
\frac{\mu^2}{ 1 + \mu^2 } \, p_T^2 \right)^2 - 2 \, H_a^2 \left( 2
+ \mu^2 \right) \left[ 2 \left( 1 + \mu^2 \right) p^2 + \frac{
\mu^2 }{ 1 + \mu^2 } \, p_T^2 \right] \right\}
\label{in-nondyn}
\end{eqnarray}

The initial conditions for $\hat\Psi, \, \dot{\hat\Psi}, \, \hat{\alpha}, \, \dot{\hat\alpha}$ are obtained from the early time expansion of the action (\ref{genYact}). Namely, we need to expand at early times the three matrices $K ,\, X ,\, \Omega^2$ entering in this action.~\footnote{The only time dependent quantities entering in these expressions are the scale factors. Some care needs to be taken in the expansion, since $H_b > H_a$, so that $p_T \gg p_L$ in the asymptotic past. The expansion (\ref{action-acw-2dS-can}) is valid as long as $H_b < 2 \, H_a \,$. If this is not the case, a term proportional to $p_L^2$ becomes smaller than a term proportional to $p_T$
at asymptotically early times, and the expansion in powers of $\left( H / p \right)$ that we have performed
(where $p$ is either of $p_L$ or $p_T$) becomes invalid. The condition $H_b < 2 \, H_a \,$ is not particularly stringent, since such a large anisotropic is certainly incompatible with the observations.} We find
\begin{equation}
S_{{\rm 2dS}} = \frac{1}{2}\, \int d^3k\, dt\, \left\{ \vert
\dot{H}_{+} \vert^2 - p^2\, \vert H_{+} \vert^2 + \vert
\dot{\Delta}_{+} \vert^2 - p^2\, \vert \Delta_{+} \vert^2 +
\frac{\mu\, H_a}{\sqrt{2}}\, \frac{p_T}{p}\, \left(
\dot{H}_{+}^*\, \Delta_{+} - \dot{\Delta}_{+}\, H_{+}^* + {\rm
h.c.} \right) \right\} \label{action-acw-2dS-can}
\end{equation}
where
\begin{equation}
\hat{\Psi} \equiv \frac{1}{\sqrt{a}\, b}\, \frac{2 p^2 + \mu^2\,
\left( p^2 + p_L^2 \right)}{\sqrt{2}\, \left(1 + \mu^2 \right)\,
p_T^2}\, H_{+} \,\,\,\,\, , \,\,\,\,\, \hat\alpha \equiv
\frac{1}{\sqrt{a}\, b}\, \left[ - \frac{p}{p_T}\, \Delta_{+} +
\frac{\mu\, \left( p^2 + p_L^2 \right)}{\sqrt{2}\, p_T^2}\, H_{+}
\right] \label{can-acw}
\end{equation}
The next to leading order corrections to (\ref{action-acw-2dS-can}) and (\ref{can-acw}) 
are of ${\rm O}\left(H_a^2/p^2\right)$ (either $p_L$ or $p_T$). Not surprisingly, the early time
frequency is controlled by the momentum term (which is a common
result in the theory of cosmological perturbations on a
inflationary background). What is nontrivial is the relation
(\ref{can-acw}) between the canonical variables and the original
modes, which, as we shall now see, is needed in order to set the
initial conditions for the latter.

Since the early time frequency varies adiabatically as
\begin{equation}
\omega^2 \approx p^2 \,\,\,\, , \,\,\,\,\,
\frac{\dot\omega}{\omega^2} \approx - \frac{p^2 + \mu^2\,
p_T^2}{p^2}\, \frac{H_a}{p} \ll 1 \nonumber
\end{equation}
we can set the adiabatic initial conditions for the canonical
variables \footnote{This is precisely what is done in the standard 
case \cite{mfb}, where
the canonical variable for the scalar perturbation is the Mukhanov-Sasaki 
\cite{musa} variable $v \,$.}
\begin{equation}
H_{+, in} = \Delta_{+,in} = \frac{1}{\sqrt{2 p}} \,\,\, , \,\,\,
\dot{H}_{+,in} = -i\, p\, H_{+,in} \,\,\, , \,\,\,
\dot{\Delta}_{+,in} = -i\, p\, \Delta_{+,in}
\end{equation}
which are ${\rm O}\left( H_a/p \right)$ accurate. Using
(\ref{can-acw}) we then obtain the initial conditions for the
original modes as
\begin{eqnarray}
\hat{\Psi}_{in} &=& \frac{1}{\sqrt{a}\, b}\, \frac{2 p^2 + \left(
p^2 + p_L^2\right)\, \mu^2}{2\, \sqrt{p}\, p_T^2\,
\left(1+\mu^2\right)}\, \left[ 1 + {\rm
O}\left(\frac{H_a^2}{p^2}\right) \right] \,\,\,\, , \,\,\,\,
\dot{\hat\Psi}_{in} = -i\, p\, \hat{\Psi}_{in}\, \left[ 1 + {\rm
O}\left( \frac{H_a}{p} \right) \right] \nonumber\\
\hat{\alpha}_{in} &=& -\frac{1}{\sqrt{a}\, b}\, \frac{\sqrt{2}\,
p\, p_T - \left( p^2 + p_L^2 \right)\, \mu}{2\, \sqrt{p}\,
p_T^2}\, \left[ 1 + {\rm O}\left(\frac{H_a^2}{p^2}\right) \right]
\,\,\,\, ,\,\,\,\, \dot{\hat\alpha}_{in} = -i\, p\,
\hat{\alpha}_{in}\, \left[ 1 + {\rm O}\left( \frac{H_a}{p} \right)
\right] \nonumber\\ \label{initialcond}
\end{eqnarray}
The initial conditions for the nondynamical modes are then obtained 
by inserting these expressions into (\ref{in-nondyn}). These are the initial conditions
for the numerical evolution discussed in the main text, which leads to the solutions given
in Figure \ref{fig:fig1}.

\end{document}